\documentclass[journal]{IEEEtran}

\usepackage{indentfirst}
\usepackage{graphicx}
\usepackage{amsmath}
\usepackage{amssymb}
\usepackage{amsfonts}
\usepackage{mathrsfs}
\usepackage{leftidx}
\usepackage{color}
\usepackage{amsmath}
\usepackage{arydshln}
\usepackage{amsthm}
\usepackage{ragged2e}
\usepackage{cite}
\usepackage{enumerate}
\usepackage{longtable}
\usepackage{float}
\usepackage{stfloats}
\usepackage{hyperref }
\usepackage{algpseudocode}
\usepackage{algorithm}
\usepackage[caption=false,font=normalsize,labelfont=sf,textfont=sf]{subfig}
\usepackage{tabularx}

\theoremstyle{plain}

\newtheorem{lem}{Lemma}
\newtheorem{prop}{Proposition}

\theoremstyle{definition}
\newtheorem{defn}{Definition}

\newtheorem{rem}{Remark}

\usepackage[table,usenames,dvipsnames]{xcolor}
\definecolor{aa}{RGB}{175,238,238}
\definecolor{bb}{RGB}{255,255,255}

\usepackage{bm}
\usepackage{makecell}

\usepackage{siunitx}

\begin{document}

\title{Efficient Gaussian Process Classification-based Physical-Layer Authentication with Configurable Fingerprints for 6G-Enabled IoT}
\author{Rui Meng, Fangzhou Zhu, Xiqi Cheng, Xiaodong Xu,~\IEEEmembership{Senior Member,~IEEE,} Bizhu Wang, Chen Dong,

Bingxuan Xu, Xiaofeng Tao,~\IEEEmembership{Senior Member,~IEEE,} and Ping Zhang,~\IEEEmembership{Fellow,~IEEE}
\thanks{The work presented in this paper was supported by the National Key R\&D Program of China No. 2020YFB1806905, the National Natural Science Foundation of China No. 61871045 and No. 61932005, Beijing Natural Science Foundation No. L242012, and the research foundation of Ministry of Education-China Mobile under Grant MCM20180101.
Corresponding author: Xiaodong Xu.}

\thanks{Rui Meng and Xiaofeng Tao are with the National Engineering Laboratory for Mobile Network Technologies, Beijing University of Posts and Telecommunications, Beijing 100876, China (email: buptmengrui@bupt.edu.cn; taoxf@bupt.edu.cn).}

\thanks{Fangzhou Zhu was with the State Key Laboratory of Networking and Switching Technology, Beijing University of Posts and Telecommunications, Beijing 100876, China. He is currently employed at the headquarters of Bank of China Limited, Beijing. (e-mail:  zfzgps7935371\_bj@bank-of-china.com).}

\thanks{Xiqi Cheng, Bizhu Wang, Bingxuan Xu, Chen Dong, and Ping Zhang are with the State Key Laboratory of Networking and Switching Technology, Beijing University of Posts and Telecommunications, Beijing 100876, China (e-mail: chengzi@bupt.edu.cn; wangbizhu\_7@bupt.edu.cn; xubingxuan@bupt.edu.cn; dongchen@bupt.edu.cn; pzhang@bupt.edu.cn).}

\thanks{Xiaodong Xu are with the State Key Laboratory of Networking and Switching Technology, Beijing University of Posts and Telecommunications, Beijing 100876, China, and also with the Department of Broadband Communication, Peng Cheng Laboratory, Shenzhen 518066, Guangdong, China (e-mail: xuxiaodong@bupt.edu.cn).}
}

\maketitle

\begin{abstract}
The future 6G-enabled IoT will facilitate seamless global connectivity among ubiquitous wireless devices, but this advancement also introduces heightened security risks such as spoofing attacks. Physical-Layer Authentication (PLA) has emerged as a promising, inherently secure, and energy-efficient technique for authenticating IoT terminals. Nonetheless, the direct application of state-of-the-art PLA schemes to 6G-enabled IoT encounters two major hurdles: inaccurate channel fingerprints and the inefficient utilization of prior fingerprint information. To tackle these challenges, we leverage Reconfigurable Intelligent Surfaces (RISs) to enhance fingerprint accuracy. Additionally, we integrate active learning and Gaussian Processes (GPs) to propose an Efficient Gaussian Process Classification (EGPC)-based PLA scheme, aiming for reliable and lightweight authentication. Following Bayes' theorem, we model configurable fingerprints using GPs and employ the expectation propagation method to identify unknown fingerprints. Given the difficulty of obtaining sufficient labeled fingerprint samples to train PLA models, we propose three fingerprint selection algorithms. These algorithms select unlabeled fingerprints and query their identities using upper-layer authentication mechanisms. Among these methods, the optimal algorithm reduces the number of training fingerprints needed through importance sampling and eliminates the requirement for PLA model retraining through joint distribution calculation. Simulations results reveal that, in comparison with non-RIS-based approaches, the RIS-aided PLA framework decreases the authentication error rate by 98.69\%. In addition, our designed fingerprint selection algorithms achieve a reduction in the authentication error rate of up to 86.93\% compared to baseline active learning schemes.
\end{abstract}


\section{Introduction}
The Internet of Things (IoT), a vast and expanded network system anchored on the Internet, employs diverse devices and technologies, including sensor equipment, GPS systems, infrared sensors, and laser scanners. These enable intelligent perception, recognition, and management of objects and machines. To achieve IoT's ultimate goal of enabling real-time interaction among everything, 6G promises enhanced services beyond those of 5G. These enhancements encompass seamless global coverage, higher data rates, and inherent security upgrades. The seamless integration of IoT facilitated by 6G necessitates a heightened involvement of radio equipment in communication processes \cite{zeng2024tutorial,dhiman2024smose}.

However, this broadening of application fields introduces more pressing security risks \cite{trevlakis2023localization}. Furthermore, due to the open broadcast nature of wireless transmission media, more attackers can forge identification information to pose as legitimate users and further manipulate their privacy data \cite{xia2021multiple}. Therefore, to achieve seamless and reliable communication, efficient and lightweight identity recognition is essential \cite{garzon2022decentralized}. Presently, in 5G systems, terminal identity identification is primarily handled by the core network, utilizing cryptographic-based authentication protocols such as 5G AKA \cite{3gpp2020security}. Despite these cryptographic techniques at upper layers being integrated into the 5G standard, they may fall short of meeting desired performance in different rising applications integrating 6G and IoT due to limited security levels \cite{xie2022security}, low compatibility among heterogeneous systems \cite{abdulqadder2022sliceblock}, and unacceptable computational overhead for resource-constrained IoT terminals \cite{fang2023collaborative}.

Consequently, to guarantee highly reliable communications in 6G-enabled IoT systems, more efficient and dependable terminal access authentication methods are imperative. Recently, Physical-Layer Authentication (PLA) has emerged as a supplementary measure to traditional security mechanisms, offering several notable advantages:
\begin{itemize}
    \item Firstly, different from the “patch” and “plugin” cryptography technologies employed at upper-layers \cite{jin2021introduction}, PLA leverages physical-layer features derived from communication links, devices, and location-related attributes, including channel fingerprints \cite{xia2021multiple,fang2023collaborative,jing2023multi,meng2023multiuser} and RF fingerprints \cite{oligeri2022past}. 
    These physical quantities serve as inherent fingerprints, characterized by time-varying, randomness, and space independence within wireless channels\cite{jin2021introduction,meng2023multiuser}. In this way, endogenous fingerprints at the physical-layer provide unique identification features and persistent defense for legitimate IoT terminals, making it extremely challenging for adversaries to extract and forge these channel characteristics \cite{xie2020survey}.
    \item Furthermore, PLA is a lightweight approach that bypasses upper-layer signaling procedures, thereby enhancing efficiency. During the channel estimation phase, the access point acquires CSI information for all franchised devices, further reducing computation cost. This allows IoT equipment with insufficient computing and storage ability to operate optimally \cite{xie2022security}.
    \item Additionally, PLA showcases robust compatibility in heterogeneous scenarios, where even equipment that can't interpret each other's upper-layer signals can still communicate via physical-layer bit-streams. Leveraging these physical-layer attributes for identification, PLA provides a seamless user experience by enabling adaptable, standardized authentication tailored to individual needs \cite{nguyen2021security}.

\end{itemize}

In earlier research, PLA was modeled as a statistical hypothesis testing process, where fingerprints were deemed as legal if the differences between them and the reference fingerprint did not surpass the detection threshold \cite{xiao2008using}. Nevertheless, accurately modeling the fingerprint distribution of IoT terminals in 6G systems is challenging due to unknown and uncertain dynamic variations, making it difficult to obtain the theoretically optimal threshold \cite{liu2022online}. Recently, Machine Learning (ML)-based PLA approaches have garnered increased interest, as intelligent algorithms possess robust feature learning abilities that enhance authentication reliability and robustness \cite{chorti2022context,meng2024survey,wang2024knowledge}. Various ML techniques, such as support vector machine (SVM) \cite{wang2021channel}, ensemble learning \cite{xie2021weighted}, deep neural networks (DNN) \cite{chen2021physical,meng2023multiuser}, and reinforcement learning (RL) \cite{xiao2016phy}, are employed to distinguish between legitimate and illegal fingerprints.
Intelligent PLA technologies can be utilized in one of six usage scenarios of 6G recommended by International Telecommunication Union (ITU): artificial intelligence and communication \cite{recommendation2023framework}. Nevertheless, there are two challenges of applying existing ML-based PLA schemes directly to 6G-empowered IoT:
\begin{itemize}
\item 

The first challenge is the inaccurate channel fingerprints within complex wireless environments. These fingerprints serve as crucial identifiers for IoT terminals, requiring high precision to exhibit distinct features essential for authentication. However, in low Signal-to-Noise Ratio (SNR) scenarios influenced by electromagnetic interference, lengthy transmission paths, and multipath effects, inaccurate fingerprint estimation significantly impairs identification performance \cite{nguyen2021security}.
While multi-attribute \cite{fang2018learning,xia2021multiple,zhang2020physical} and multi-observation \cite{senigagliesi2022authentication,xie2021weighted,xiao2017phy,meng2024multiobservation} approaches enhance the distinguishability of legitimate IoT terminals, the fundamental challenge of fingerprint estimation remains unresolved. In fact, these approaches may introduce new issues, such as increased deployment costs and security considerations associated with multi-receiver setups.

\item 
The second challenge involves the inefficient utilization of prior information of fingerprints. Supervised learning-based PLA schemes depend expert databases of fingerprints to train models effectively. However, obtaining identity labels through upper-layer authentication mechanisms escalates computational costs as the number of training samples increases. Moreover, not all training samples contribute equally to feature learning, extraction, and analysis; some may be redundant or trivial. Although semi-supervised learning approaches like variational autoencoders \cite{meng2023physical}, clustering \cite{xia2021multiple}, and one-class classifiers \cite{senigagliesi2022authentication,wang2021channel} circumvent the need for attacker-specific prior information, they face limitations in detecting attacks due to the absence of labeled attacker fingerprints for guiding model training.

\end{itemize}

In response to these challenges, we utilize Reconfigurable Intelligent Surfaces (RISs), also known as Intelligent Reflecting Surfaces (IRSs), to establish an alternative transmission path that offers improved channel quality, thereby enhancing the reliability of channel fingerprints. RISs represent a groundbreaking technology in 6G, utilizing a planar array of cost-effecient reflective elements to vibrantly control amplitudes, phases, and frequencies of signals \cite{chen2023optimal}. This capability significantly enhances overall communication performance \cite{chen2022accurate}. RISs are primarily controlled by base stations (BSs) to extend coverage, especially in high-frequency bands such as millimeter waves. Importantly, these deployments also hold promise for authentication purposes. For example, Gao et al. \cite{gao2024physical} combine RISs and key-based authentication approaches to improve the key generation rate. Furthermore, we leverage Gaussian processes (GPs) \cite{williams2006gaussian} to model configurable fingerprints introduced by RISs for the following reasons: GPs possess the flexibility to model complex and nonlinear relationships and offer a probabilistic depiction of model uncertainty. Given that various optimization approaches \cite{zhu2022intelligent} can dynamically adjust the parameters of RISs for resource allocation and energy efficient maximization, the statistical characteristics of channels vary over time and space. GPs are adept at capturing these variations and handling uncertainties, thereby effectively fitting the fingerprint distribution. For instance, Wang et al. \cite{wang2021channel} utilize GP regression to predict mobile transmitters' CSI fingerprints. Moreover, to efficiently utilize prior fingerprint information for guiding the training of PLA models, we employ active learning \cite{settles2009active} to facilitate interaction between PLA and upper-layer authentication mechanisms. The fingerprint selection algorithm chooses representative unlabeled fingerprints, which are then tagged with identity labels through upper-layer authentication mechanism, thereby guiding the training of the PLA model. This iterative process enables the achievement of robust PLA models with a minimal set of labeled training fingerprints. To summarize briefly, the main contributions are as below.

\begin{itemize}
\item We introduce an Efficient Gaussian Process Classification (EGPC)-based PLA scheme for 6G-enabled IoT to realize lightweight and reliable cross-layer identity recognition. RISs are utilized to obtain configurable fingerprints with enhanced accuracy. Active learning is employed to selectively label these fingerprints, thereby maximizing the utilization efficiency of existing fingerprint data.
\item Configurable fingerprints are modeled using GPs in our proposed Gaussian Process Classification (GPC)-based PLA approach. Additionally, the expectation propagation algorithm is utilized to approximate the posterior prediction of unknown fingerprint identities.
\item We present three fingerprint selection algorithms, including Random Optimization (RO), Average Loss of Uncertainty (ALU), and Soft-ALU (SALU). These algorithms prioritize the labeling of the most uncertain fingerprint sample from the unlabeled pool. The design enhances interaction efficiency between PLA and upper-layer authentication mechanisms.
\item Simulations conducted on synthetic dataset demonstrate the efficacy of the designed scheme. Moreover, our algorithms outperform three baseline active learning schemes in terms of authentication error rate.
\end{itemize}


\section{System Model and Problem Formulation}
\label{model}
\subsection{Network Model}

\begin{figure}[h]
\centering
\includegraphics[width=0.5\textwidth]{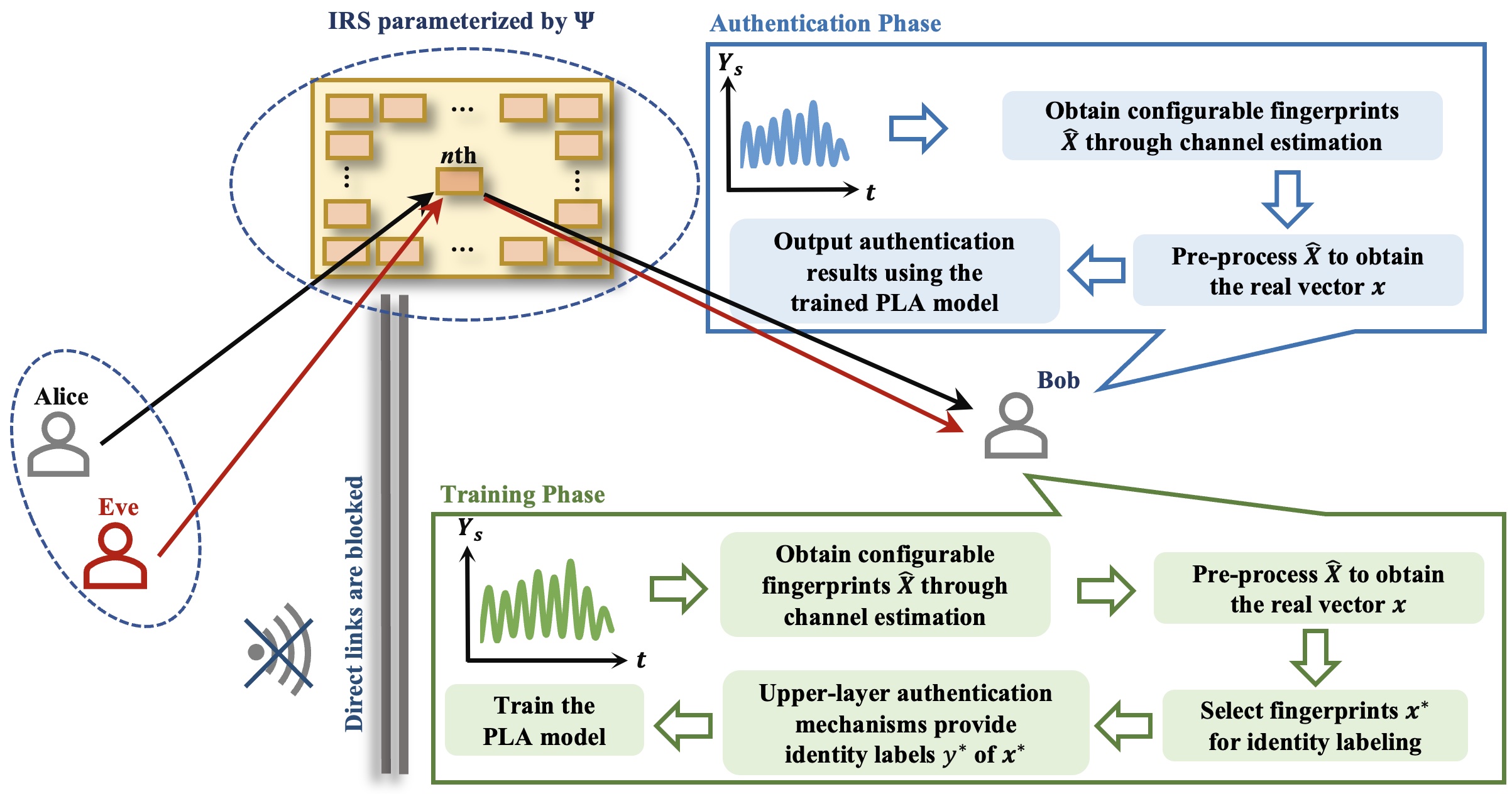}
\caption{The designed EGPC-based PLA scheme, where Alice and Eves represent legitimate transmitters and spoofers, respectively, and Bob is responsible for identity recognition.}
\label{fig1}
\vspace{-1em}
\end{figure}

As shown in Fig. \ref{fig1}, a standard security model is considered in this study, which is described in detail as follows.
\begin{itemize}
    \item \textit{Alice (legal IoT terminal):} This paper concentrates on the authentication of uplink channel signals. Alice transmits signals to Bob to initiate identity authentication requests.
    \item \textit{Eve (spoofing attacker):} Eve, acting as a spoofing attacker, attempts to impersonate Alice by transmitting signals to Bob during different communication time slots. To ensure a realistic scenario, Eve is assumed to be positioned more than half a wavelength away from Alice, which guarantees strong spatial decorrelation between their channel fingerprints.
    \item \textit{Bob (BS):} Bob, the BS, is responsible for identifying the identities of the received signals. The ML-based PLA model is trained and deployed at Bob. To enhance efficiency, the training process can leverage edge computing assistance, enabling faster and more resource-effective model optimization.
    \item \textit{RIS:} The accuracy and reliability of fingerprints in low SNR environments are inherently limited, thus impacting the performance of PLA methods. In such scenarios, RISs can facilitate wireless connectivity between Alice and Bob. RIS intelligently reflects wireless signals to extend the coverage area, reduce signal blind spots, and enhance signal strength and quality, thereby improving the accuracy of fingerprints. Additionally, RIS can dynamically adjust the phase and amplitude of its reflective elements according to environmental changes, thereby mitigating the impact of multipath effects and interference. Through configurable adjustments to the RIS settings, Bob can effectively manipulate the channel between Alice and himself, thereby bolstering the spatial discernibility of fingerprints. The integration of RISs within authentication protocols presents significant potential, particularly within controlled settings such as smart factories \cite{tomasin2022challenge}. Incorporating RISs within authentication procedures enhances security measures, ensuring robust identity verification and access control \cite{wang2022wireless}. The absence of direct links between transmitters and Bob is assumed for simplicity.
\end{itemize}

\begin{table}[htb] 
\begin{center}   
\caption{The List of Main Parameters} 
\setlength{\tabcolsep}{4pt}

\label{table1} 
\renewcommand{\arraystretch}{1.2} 
\begin{tabular}{|c|c|}  
\hline   \textbf{Notations} & \textbf{Meaning} \\   
\hline $\mathbb{E}$ & Expectation  \\
\hline $\mathbb{V}$ & Variance  \\
\hline   $\propto$ & Proportional to  \\      
\hline   $\sim$ & Distributed according to \\
\hline   $\bm{c}^T$ & The transpose of vector $\bm{c}$  \\
\hline $y$ & The identity label \\
\hline $\sigma(z)$ & The sigmoid function \\
\hline   $\triangleq$ & An equality which acts as a definition  \\  
\hline   $\bm{f}_\ast$ & Gaussian process posterior prediction \\
\hline   $\bm{X}$ & Signal at the transmitter \\
\hline   $\bm{Y}_S$ & Signal at the receiver \\
\hline   $\bm{\Psi}$ & The RIS element response matrix \\
\hline $\bm{c}$ & The estimated configurable fingerprints \\
\hline $f(\bm{c})$ or $\bm{f}$ & Gaussian process latent function values \\
\hline ${\bar{\pi}}_\ast$ & The probabilistic prediction of the identity \\
\hline $Z$ & The marginal likelihood \\
\hline $\Phi\left(z\right)$ & The CDF of a standard normal distribution \\
\hline $t_l\left(f_l\middle|{\widetilde{Z}}_l,{\widetilde{\mu}}_l,{\widetilde{\sigma}}_l^2\right)$ & The local likelihood approximation \\
\hline $q\left(\bm{f}\middle| C,\bm{y}\right)$ & \makecell{The distribution posterior used \\ to approximate $p\left(\bm{f}\middle| C,\bm{y}\right)$ }  \\
\hline $K$ & The covariance (or Gram) matrix \\
\hline $\bm{k}_\ast$ & Short for $ K\left(\bm{c},\bm{c}_\ast\right)$ \\
\hline $q_{-l}\left(f_l\right)$ & The cavity distribution \\
\hline
\end{tabular}   
\end{center}   
\end{table}

\subsection{Channel Model}
Let $N_T$ and $N_R$ denote the number of antennas for Alice and Bob, respectively. The received signal vector at Bob, denoted as an $N_R$-size column vector $\bm{Y}_S$, can be formulated as
\begin{equation}
\label{signal}
\bm{Y}_S=\bm{Q}^{\left(A,I,B\right)}\bm{X}+\bm{W},
\end{equation}
where $\bm{X}$ is the transmitted signal vector from Alice with an $N_T$-size column, and $\bm{W} \sim \mathcal{CN}(0,\bm{\sigma}^2)$ is the Gaussian noise vector with an $N_R$-size column. Let $N$ be the number of elements in the RIS. The cascade channel matrix from Alice to Bob via the RIS is given by $\bm{Q}^{\left(A,I,B\right)}=\bm{Q}^{(R,B)}\bm{\Psi}\bm{Q}^{(A,R)}$, where $\bm{Q}^{(R,B)}$ and $\bm{Q}^{(A,R)}$ are the $N_R\times N$ channel matrix from the RIS to Bob and the $N\times N_T$  channel matrix from Alice to the RIS, respectively. The matrix $\bm{\Psi}=diag\left(\psi_0,\ldots,\psi_{N-1}\right)$ represents the responses of the RIS elements. Each element $\psi_n=A_n\left(\theta_n\right)e^{j\theta_n}$ is the response of the $n$th RIS element, where $A_n\left(\theta_n\right)$ and $e^{j\theta_n}$ denote its controllable magnitude and phase response, respectively. Configurable fingerprints $\bm{c}$ are obtained through
\begin{equation}
{\bm{c}}=f_{ce}(\bm{Y}_S),
\end{equation}
where $f_{ce}$ denotes the channel estimation function. Notably, we do not focus on the channel estimation, which can be achieved through various approaches, such as classical channel estimation, compressed sensing, matrix factorization, and deep learning approaches \cite{zheng2022survey,ghosh2024ee}.


\subsection{Problem Formulation}
\label{problem}
We preprocess $\bm{c}$ to obtain a $2N_RN_T$-dimensional real configurable fingerprint vector $\bm{c}$. Let $\mathcal{C}$ represent the configurable fingerprint space, and $\mathcal{Y}$ denote the set of identity labels. Our goal is to obtain the trained authenticator, denoted as $\psi:\ \mathcal{C}\rightarrow\mathcal{Y}$, which predicts the identities of $\psi(\bm{c}_\ast)$ for $\bm{c}_\ast\in\mathcal{C}$. This will be further discussed in Sec. \ref{method}.

Given that RISs introduce a more intricate signal processing framework, and that the identity labeling of fingerprints requires involvement in the upper authentication protocol, it becomes crucial to devise an efficient fingerprint labeling strategy. This strategy is formulated as
\begin{equation}
f_{fl}=(\mathcal{U},\mathcal{L}),
\end{equation}
where $\mathcal{U}$ and $\mathcal{L}$ represent unlabeled and labeled fingerprints, respectively, and $f_{fl}$ is the fingerprint labeling function. $f_{fl}$ outputs the selected unlabeled fingerprints $\bm{c}^\ast\in\mathcal{U}$ for identity labeling. Through iterative selection and labeling, PLA scheme's dependability is progressively enhanced. The fingerprint labeling strategy will be discussed in Sec. \ref{method2}. The main parameters are illustrated in Tab. \ref{table1}.



\section{Posterior Prediction of Unknown Configurable Fingerprints}
\label{method}

GP is a non-parametric probabilistic model that can dynamically adapt to complex and varying channel conditions. It effectively captures and simulates the nonlinear characteristics of low SNR environments through kernel functions, without requiring the assumption of a specific parametric model structure. Additionally, GP can dynamically adjust its model parameters based on new channel conditions through its Bayesian framework, adeptly handling the inherent uncertainties in channel characteristics across time, space, and frequency domains, making it an ideal choice for fitting configurable fingerprint distributions.
Hence, this section introduces a GPC-based PLA model to derive posterior predictions of fingerprints.

\subsection{Gaussian Process (GP) for Configurable Fingerprints}

As mentioned in Sec. \ref{problem}, we formulate the problem as a classification problem, denoted by $\psi$, which assigns $\bm{c}$ to one of two identity labels: $y=0$ for Alice and $y=1$ for Eve.
We utilize a GPC framework to establish a connection between these configurable fingerprints and their corresponding identity classes through a latent function $f$, which follows a GP prior $f \sim GP(\mu\left(\cdot\right),k(\cdot,\cdot))$, where $\mu\left(\cdot\right)$ represents the mean function and $k(\cdot,\cdot)$ represents the covariance kernel function.

Through “squashing” $f(\bm{c})$ through the logistic function, we can get a prior on $\pi(\bm{c}) p (y=+1 \mid \bm{c})=\sigma(f(\bm{c}))$. While we are primarily interested in $\pi\left(\bm{c}\right)$, the function $f(\bm{c})$ facilitates the formulation of our model. To eliminate $f(\bm{c})$ and compute the distribution of the latent variable for unknown configurable fingerprints, we use the following equation:
\begin{equation}
\label{6}
p\left(f_\ast\middle| C,\bm{y},\bm{c}_\ast\right)=\int{p\left(f_\ast\middle| C,\bm{c}_\ast,\bm{f}\right)p\left(\bm{f}\middle| C,\bm{y}\right)d\bm{f}},
\end{equation}
where $p\left(\bm{f}\middle| C,\bm{y}\right)=p\left(\bm{y}\middle|\bm{f}\right)p\left(\bm{f}\middle| C\right)/p(\bm{y}|C)$ represents the posterior distribution over the latent variables. Subsequently, $p\left(f_\ast\middle| C,\bm{y},\bm{c}_\ast\right)$ is utilized to get the probabilistic prediction of identities:
\begin{equation}
\label{7}
{\bar{\pi}}_\ast \triangleq p\left(y_\ast=+1\middle| C,\bm{y},\bm{c}_\ast\right)=\int{\sigma\left(f_\ast\right)p\left(f_\ast\middle| C,\bm{y},\bm{c}_\ast\right)df_\ast}.
\end{equation}
The integrals in (\ref{6}) and (\ref{7}) are analytically intractable due to the non-Gaussian likelihood. Therefore, we employ the expectation propagation method \cite{williams2006gaussian} to approximate the non-Gaussian joint posterior.

\subsection{Expectation Propagation-based Posterior Prediction}
Based on Bayes’ theorem, the posterior distribution $p\left(\bm{f}\middle| C,\bm{y}\right)$ is given by
\begin{equation}
\label{8}
p\left(\bm{f}\middle| C,\bm{y}\right)=\frac{1}{Z}p\left(\bm{f}\middle| C\right)\prod_{l=1}^{n}{p\left(y_l\middle| f_l\right)},
\end{equation}
where $p\left(\bm{f}\middle| C\right)$ represents the Gaussian prior, $\prod_{l=1}^{n}{p\left(y_l\middle| f_l\right)}$ denotes the likelihood, and $n$ is the total number of training configurable fingerprints. The marginal likelihood is given by
\begin{equation}
\label{9}
Z=p\left(\bm{y}\middle| C\right)=\int{p\left(\bm{f}\middle| C\right)\prod_{l=1}^{n}{p\left(y_l\middle| f_l\right)}d\bm{f}}.
\end{equation}
To render $p\left(\bm{f}\middle| C,\bm{y}\right)$ analytically intractable, we employ the probit likelihood:
\begin{equation}
p\left(y_l\middle| f_l\right)=\Phi\left(f_ly_l\right),
\end{equation}
where $\Phi\left(z\right)=\int_{-\infty}^{z}{\mathcal{N}(c|0,1)dc}$ denotes the cumulative density function (CDF) of the standard Gaussian distribution. The likelihood $\Phi\left(f_ly_l\right)$ is approximated using the local likelihood approximation:
\begin{equation}
\label{11}
p\left(y_l\middle| f_l\right) \simeq t_l\left(f_l\middle|{\widetilde{Z}}_l,{\widetilde{\mu}}_l,{\widetilde{\sigma}}_l^2\right)={\widetilde{Z}}_l\mathcal{N}\left(f_l\middle|{\widetilde{\mu}}_l,{\widetilde{\sigma}}_l^2\right),
\end{equation}
where ${\widetilde{Z}}_l$, $ {\widetilde{\mu}}_l$, and $ {\widetilde{\sigma}}_l^2$ are site parameters, and the local likelihoods $t_l$ satisfy
\begin{equation}
\prod_{l=1}^{n}{t_l\left(f_l\middle|{\widetilde{Z}}_l,{\widetilde{\mu}}_l,{\widetilde{\sigma}}_l^2\right)}=\mathcal{N}\left(\widetilde{\bm{\mu}},\widetilde{\Sigma}\right)\prod{\widetilde{Z}}_l,
\end{equation}
where $\widetilde{\bm{\mu}}$ is the vector of ${\widetilde{\mu}}_l$ and $\widetilde{\Sigma}$ is diagonal with ${\widetilde{\Sigma}}_{ii}={\widetilde{\sigma}}_l^2$. 

We can approximate the posterior of $p\left(\bm{f}\middle| C,\bm{y}\right)$ as
\begin{equation}
\label{13}
q\left(\bm{f}\middle| C,\bm{y}\right)\triangleq \frac{1}{Z_{EP}}p\left(\bm{f}\middle| C\right)\prod_{l=1}^{n}{t_l\left(f_l\middle|{\widetilde{Z}}_l,{\widetilde{\mu}}_l,{\widetilde{\sigma}}_l^2\right)}=\mathcal{N}\left(\bm{\mu},\Sigma\right),
\end{equation}
where $\bm{\mu}=\Sigma{\widetilde{\Sigma}}^{-1}\widetilde{\bm{\mu}}$ and $\Sigma={(K^{-1}+{\widetilde{\Sigma}}^{-1})}^{-1}$. Here, $K$ is a $n\times n$ covariance matrix, and $Z_{EP}=q(\bm{y}|C)$ represents the approximation to $Z$ from (\ref{8}) and (\ref{9}).

\begin{rem}
One viable approach to selecting the parameters of $t_l$ is to minimize the reverse Kullback-Leibler (KL) divergence $KL(q\left(\bm{f}\middle| C,\bm{y}\right)||p\left(\bm{f}\middle| C,\bm{y}\right))$ with respect to $q\left(\bm{f}\middle| C,\bm{y}\right)$, a process known as variational inference. In this paper, we utilize the expectation propagation algorithm to sequentially obtain the proximity of $t_l$ as discussed below.
\end{rem}

\subsubsection{Marginal Cavity Distribution}
By combining $p\left(\bm{f}\middle| C\right)$ with $\prod_{l=1}^{n}{t_l\left(f_l\middle|{\widetilde{Z}}_l,{\widetilde{\mu}}_l,{\widetilde{\sigma}}_l^2\right)}$, the cavity distribution is derived as follows:
\begin{equation}
\label{14}
q_{-l}\left(f_l\right)\propto\int{p\left(\bm{f}\middle| C\right)\prod_{j\neq i}{t_j\left(f_j\middle|{\widetilde{Z}}_j,{\widetilde{\mu}}_j,{\widetilde{\sigma}}_j^2\right)}df_j}.
\end{equation}
We combine $p\left(\bm{f}\middle| C\right)$ and the $(n-1)$ approximate likelihoods in (\ref{14}) through equivalently removing $t_l\left(f_l\middle|{\widetilde{Z}}_l,{\widetilde{\mu}}_l,{\widetilde{\sigma}}_l^2\right)$ from the approximated posterior $\prod_{l=1}^{n}{t_l\left(f_l\middle|{\widetilde{Z}}_l,{\widetilde{\mu}}_l,{\widetilde{\sigma}}_l^2\right)}$ in (\ref{13}).

\begin{prop}
\label{prop2}
We can obtain the marginal distribution for $f_l$ from $q\left(\bm{f}\middle| C,\bm{y}\right)$ as
\begin{equation}
\label{15}
q\left(f_l\middle| C,\bm{y}\right)=\mathcal{N}\left(f_l\middle|\mu_l,\sigma_l^2\right),
\end{equation}
where $\sigma_l^2=\Sigma_{ii}$.
\end{prop}

\begin{IEEEproof}
    See Appendix \ref{appa}.  
\end{IEEEproof}

\begin{prop}
\label{prop3}
By dividing (\ref{15}) by $t_l$, we can derive the cavity distribution as
\begin{equation}
q_{-l}\left(f_l\right)=\mathcal{N}\left(f_l\middle|\mu_{-l},\sigma_{-l}^2\right),
\end{equation}
where $\mu_{-l}=\sigma_{-l}^2(\sigma_l^{-2}\mu_l-{\widetilde{\sigma}}_l^{-2}{\widetilde{\mu}}_l)$ and $\sigma_{-l}^2={(\sigma_l^{-2}-{\widetilde{\sigma}}_l^{-2})}^{-1}$.
\end{prop}
\begin{IEEEproof}
    See Appendix \ref{appb}.  
\end{IEEEproof}

\subsubsection{The Gaussian Approximation to the Non-Gaussian Marginal Distribution}
The product of $q_{-l}\left(f_l\right)$ and $p\left(y_l\middle| f_l\right)$ can be approximated by a non-normalized Gaussian marginal distribution, defined as
\begin{equation}
\hat{q}\left(f_l\right)\triangleq{\hat{Z}}_l\mathcal{N}\left({\hat{\mu}}_l,{\hat{\sigma}}_l^2\right)\simeq q_{-l}\left(f_l\right)p\left(y_l\middle| f_l\right).
\end{equation}

\begin{prop}
\label{prop4}
Given that $\hat{q}\left(f_l\right)$ is un-normalized, to align with $q_{-l}\left(f_l\right)p\left(y_l\middle| f_l\right)$, the desired posterior marginal moments should satisfy
\begin{equation}
{\hat{Z}}_l=\Phi\left(z_l\right),
\end{equation}
\begin{equation}
{\hat{\mu}}_l=\mu_{-l}+\frac{y_l\sigma_{-l}^2\mathcal{N}\left(z_l\right)}{\Phi\left(z_l\right)\sqrt{1+\sigma_{-l}^2}},
\end{equation}
and
\begin{equation}
{\hat{\sigma}}_l^2=\sigma_{-l}^2-\frac{\sigma_{-l}^4\mathcal{N}\left(z_l\right)}{\left(1+\sigma_{-l}^2\right)\Phi\left(z_l\right)}\left(z_l+\frac{\mathcal{N}\left(z_l\right)}{\Phi\left(z_l\right)}\right),
\end{equation}
where $z_l=\frac{y_l\mu_{-l}}{\sqrt{1+\sigma_{-l}^2}}$.
\end{prop}
\begin{IEEEproof}
    See Appendix \ref{appc}.  
\end{IEEEproof}
\subsubsection{The Computation of $ t_l$}
\begin{prop}
\label{prop5}
multiplying $\hat{q}\left(f_l\right)={\hat{Z}}_l\mathcal{N}\left({\hat{\mu}}_l,{\hat{\sigma}}_l^2\right)$ by $q_{-l}\left(f_l\right)=\mathcal{N}\left(f_l\middle|\mu_{-l},\sigma_{-l}^2\right)$, the site parameters of $p\left(y_l\middle| f_l\right)\simeq{\widetilde{Z}}_l\mathcal{N}\left(f_l\middle|{\widetilde{\mu}}_l,{\widetilde{\sigma}}_l^2\right)$ in (\ref{11}) are obtained as
\begin{equation}
\label{19-1}
{\widetilde{\mu}}_l={\widetilde{\sigma}}_l^2\left({\widetilde{\sigma}}_l^{-2}{\hat{\mu}}_l-\sigma_{-l}^{-2}\mu_{-l}\right),
\end{equation}
\begin{equation}
\label{19-2}
{\widetilde{\sigma}}_l^2=\left({\hat{\sigma}}_l^{-2}-\sigma_{-l}^{-2}\right)^{-1},
\end{equation}
and
\begin{equation}
\label{19-3}
{\widetilde{Z}}_l={\hat{Z}}_l\sqrt{2\pi}\sqrt{\sigma_{-l}^2+{\widetilde{\sigma}}_l^2}exp\left(\frac{\left(\mu_{-l}-{\widetilde{\mu}}_l\right)^2}{2(\sigma_{-l}^2+{\widetilde{\sigma}}_l^2)}\right).
\end{equation}
\end{prop}

\begin{IEEEproof}
    See Appendix \ref{appd}.  
\end{IEEEproof}

By iteratively matching the moments of marginal posteriors, $p\left(\bm{f}\middle| C,\bm{y}\right)$ with the Gaussian approximation $q\left(\bm{f}\middle| C,\bm{y}\right)$ is obtained.

\begin{prop}
\label{prop6}
There is a closed representation for $p\left(y_\ast\middle| C,\bm{y},\bm{c}_\ast\right)$, which is given by
\begin{equation}
\label{20}
\begin{aligned}
& p\left(y_\ast\middle| C,\bm{y},\bm{c}_\ast\right) \\
& = \Phi\left(\frac{\bm{k}_\ast^T\left(K+\widetilde{\Sigma}\right)^{-1}\widetilde{\bm{\mu}}}{\sqrt{1+k\left(\bm{c}_\ast,\bm{c}_\ast\right)-\bm{k}_\ast^T\left(K+\widetilde{\Sigma}\right)^{-1}\bm{k}_\ast}}\right),
\end{aligned}
\end{equation}
where $\bm{k}_\ast=K(C,\bm{c}_\ast)$, $\bm{k}_\ast^T$ denotes the transpose of $\bm{k}_\ast$, and $k\left(\bm{c}_\ast,\bm{c}_\ast\right)$ is covariance functions evaluated at $\bm{c}_\ast$ and $\bm{c}_\ast$.
\end{prop}

\begin{IEEEproof}
    See Appendix \ref{appe}.  
\end{IEEEproof}
Based on $p\left(y_\ast\middle| C,\bm{y},\bm{c}_\ast\right)$, the identity of the unknown fingerprints $\bm{c}_\ast$ is obtained through ${\arg \max}_{y_\ast}{p\left(y_\ast\middle| C,\bm{y},\bm{c}_\ast\right)}$ \cite{dalton2013optimal}.

\subsection{Computational Complexity Analysis}

Computational Complexity of Covariance Matrix: In the scenario with one legitimate transmitter and one spoofing attacker, the size of the covariance matrix is $n_t\times n_t$ (where $n_t$ is the number of fingerprint samples), with a computational complexity of $O(n_t^3)$. When there are multiple legitimate transmitters and attackers ($k_t$ IoT devices), the dimension of the covariance matrix expands to $k_tn_t\times k_tn_t$, resulting in an increased computational complexity of $O((k_tn_t)^3)$. This represents a cubic polynomial growth relative to the number of IoT terminals, $k_t$.

Computational Complexity of Mean Vector: Its update complexity is much lower than that of the covariance matrix and does not change significantly with the number of IoT terminals.

\section{Proposed Efficient Gaussian Process Classification (EGPC)-based PLA Scheme}
\label{method2}

In this section, building upon the proposed GPC-based PLA scheme in Sec. \ref{method}, we present an active learning-based lightweight and efficient PLA approach. Additionally, we introduce three fingerprint selection algorithms designed to facilitate interaction between the PLA and upper-layer authentication mechanisms.

\subsection{Active Learning-based Efficient Fingerprint Labeling}

Existing ML-aided PLA approaches are divided into two major classes: supervised and semi-supervised learning-based methods. For the former, all fingerprint samples are labeled. However, acquiring sufficient labeled fingerprints in real-world wireless systems is often impractical due to attackers' concealment. This challenge is exacerbated in RIS systems with dynamic programmable characteristics, where obtaining prior information about attackers becomes even more difficult. 
Conversely, semi-supervised learning schemes involve labeling only a subset of fingerprint samples. Yet, the scarcity of prior information poses difficulties in building a dependable and resilient authentication system. Specifically, similarities and redundancies in feature patterns among labeled fingerprints not only escalate computational labeling costs but also curtail the PLA model's generalization capability. The integration of RISs further alters the fingerprint distribution, necessitating representative samples to effectively train an accurate authentication model.

\begin{figure}[t]
\centering
\includegraphics[width=0.35\textwidth]{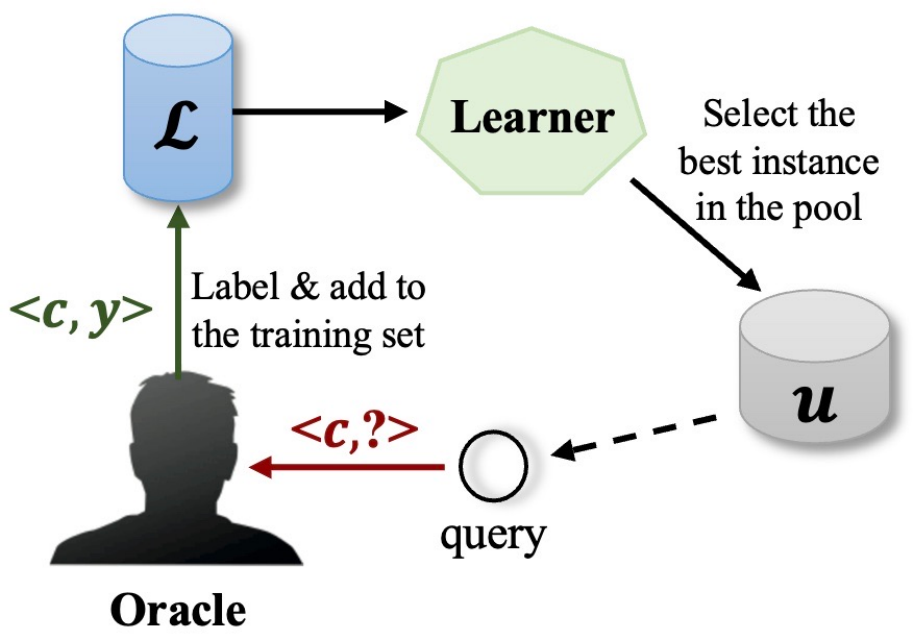}
\caption{Active learning based on pool, where the learner and oracle interact continuously to optimize its parameters.}
\label{fig2}
\vspace{-1em}
\end{figure}

To train a robust PLA model efficiently using a limited number of labeled configurable fingerprint samples, it is crucial to carefully select those samples. To achieve this, we employ active learning as our strategy. As depicted in Fig. \ref{fig2}, active learning\footnote{The paper primarily focuses on pool-based active learning, which is one of the three major scenarios in active learning. The other scenarios include membership query synthesis and stream-based selective sampling.} typically assumes the availability of a small set of labeled data, denoted as $\mathcal{L}$, and a large pool of unlabeled data, denoted as $\mathcal{U}$. The learner can submit “queries”, generally in the form of unlabeled samples, to the “oracle” for labeling, such as a human annotator. These queries are selected in a greedy manner based on a utility measure that evaluates all unlabeled samples in the pool. Active learning approaches have been researched and applied to various problems, including information extraction, video classification, and image classification, among others.

\begin{rem}
In practical applications, PLA does not serve as a replacement for upper-layer security mechanisms but rather complements them, thereby enhancing overall security \cite{xie2022security,xie2020survey}.
The proposed EGPC-based PLA scheme, inspired by active learning, involves continuous interaction with the upper-layer authentication protocol to iteratively update labeled fingerprints, as outlined in \textbf{Algorithm 1}. 
Active learning significantly reduces the reliance on a large number of labeled fingerprint samples, thereby lowering the cost and time associated with fingerprint identity labeling based on upper-layer authentication protocols. Additionally, active learning can select the most informative fingerprint samples for training, maximizing the learning effect of the PLA model within a limited training fingerprint set. Furthermore, active learning typically employs an iterative training process, where only a small number of fingerprint samples are labeled and updated each time, further reducing the consumption of computational resources.
Specifically, the fingerprint selection algorithm selects unlabeled fingerprint samples $\bm{c}_\ast$, whose identity labels are provided by upper-layer authentication mechanisms. Upon receiving new labeled fingerprint samples, the authentication model dynamically adjusts its parameters, continuing the learning process until the desired authentication performance is achieved. Therefore, this lightweight cross-layer identification scheme enhances identity recognition efficiency and bolsters security. Notably, the selection of $\bm{c}_\ast$ plays a crucial role in the PLA model's training efficiency. In Sec. \ref{selection} and Sec. \ref{selection2}, we will introduce three fingerprint selection algorithms designed to select the most uncertain fingerprint sample $\bm{c}_\ast$ from a set of unlabeled fingerprint samples $\mathcal{U}$. 
\end{rem}

\begin{algorithm}[!h]
\caption{Process of the Designed EGPC-based PLA Scheme}
\textbf{Input:} $\mathcal{U}$: ${\{\bm{c}^{(u)}\}}_{u=1}^U$; $\mathcal{L}$: ${\{{<\bm{c},y>}^{(l)}\}}_{l=1}^L$

\textbf{Process:}
\begin{algorithmic}[1]
\label{algorithm1}
\For{$t=1,2,\ldots$}
\State Training the authenticator $\theta$ by the provided GPC-based PLA scheme in Sec. \ref{method};
\State Selecting $\bm{c}^\ast\in\mathcal{U}$ according to \textbf{Algorithm 2 or 3};
\State Querying upper-layer authentication mechanisms to get the identities $y^\ast$ of fingerprints $\bm{c}^\ast$;
\State Adding $<\bm{c}^\ast,y^\ast>$ to $\mathcal{L}$;
\State Removing $\bm{c}^\ast$ from $\mathcal{U}$
\EndFor

\textbf{Output:} The trained authenticator
\end{algorithmic}
\end{algorithm}

\subsection{Proposed Random Optimization (RO)-based Fingerprint Selection Algorithm}
\label{selection}
During each iteration, we select the optimal $\bm{c}_\ast$ by optimizing the acquisition function typically involving $\pi\left(\bm{f}|C,\bm{y}\right)$. For simplicity, we omit $C$ and $\bm{y}$ from the notations, representing them as $\pi\left(\bm{f}\right)$ and $p(y|\bm{c})$ respectively. The Optimal Bayesian Classifier (OBC) \cite{dalton2013optimal} is denoted by $\psi_{\pi\left(\bm{f}\right)}(\cdot)$. The expected authentication error of the OBC can be expressed as
\begin{equation}
\begin{split}
\mathbb{E}_{\bm{c}_s}\left\{1-p\left(y_s=\psi_{\pi\left(\bm{f}\right)}\left(\bm{c}_s\right)\middle|\bm{c}_s\right)\right\} \\
=\mathbb{E}_{\bm{c}_s}\left\{1-\max_{y_s}{p\left(y_s\middle|\bm{c}_s\right)}\right\}.
\end{split}
\end{equation}

Initially, we randomly sample a fingerprint set $\mathcal{C}_\ast\subset\mathcal{C}$ of size $M_1$, calculate the acquisition function, and select the the optimal fingerprints $\bm{c}_\ast$. Let $U^A\left(\bm{c}_\ast\right)=\mathbb{E}_{\bm{c}_s}{ \{ g^A\left(\bm{c}_s;\bm{c}_\ast\right) \} }$ and $U^S\left(\bm{c}_\ast\right)=\mathbb{E}_{\bm{c}_s}{ \{ g^S\left(\bm{c}_s;\bm{c}_\ast\right) \} }$. We denote either $g^A\left(\bm{c}_s;\bm{c}_\ast\right)$ or $g^S\left(\bm{c}_s;\bm{c}_\ast\right)$ by $g\left(\bm{c}_s;\bm{c}_\ast\right)$, which can be computed given $p\left(y_s\middle|\bm{c}_s\right)$ and $p\left(y_s\middle|\bm{c}_s,\bm{c}_\ast,y_\ast\right)$. By iteratively retraining the GPC-based authenticator using the expectation propagation algorithm, the acquisition function is optimized. The steps of the proposed RO-based fingerprint selection algorithm are outlined in \textbf{Algorithm 2}.

\begin{algorithm}[!h]
\caption{Process of the designed RO-based Fingerprint Selection Algorithm}
\textbf{Input:} $p(\bm{c})$; $q(\bm{f}|C,\bm{y})$

\textbf{Process:}
\begin{algorithmic}[1]
\label{algorithm2}
\State Sampling $M_1$ fingerprints $\bm{c}_\ast \sim p(\bm{c})$ and $M_2$ fingerprints $\bm{c}_s \sim p(\bm{c})$
\For{each $\bm{c}_\ast$}
\State Computing $p\left(y_\ast\middle|\bm{c}_\ast\right)$ using (\ref{20})
\For{$y_\ast$ in $\{0,1\}$ }
\State Approximating $q\left(\bm{f}\middle|\bm{c}_\ast,y_\ast\right)$ using (\ref{19-1}), (\ref{19-2}), and (\ref{19-3})
\For{each $\bm{c}_s$}
\State Computing $p\left(y_s\middle|\bm{c}_s\right)$ and $p\left(y_s\middle|\bm{c}_s,\bm{c}_\ast,y_\ast\right)$
\State Computing $g(\bm{c}_s;\bm{c}_\ast)$
\EndFor
\EndFor
\State $U\left(\bm{c}_\ast\right)=\frac{1}{M_2}\sum_{\bm{c}_s}{g(\bm{c}_s;\bm{c}_\ast)}$ 
\EndFor

\textbf{Output:} $\widetilde{\bm{c}}=\arg \max_{\bm{c}_\ast}{U(\bm{c}_\ast)}$
\end{algorithmic}
\end{algorithm}

\subsection{Proposed Average Loss of Uncertainty (ALU)-based and Soft-ALU (SALU)-based Fingerprint Selection Algorithms}
\label{selection2}
\begin{rem}
\label{remark3}
In actual communications, \textbf{Algorithm 2} faces two issues: 1) The number of fingerprints  $M_2$ is substantial to ensure a highly reliable approximation of the integral in (\ref{22}); and 2) The computation of $p\left(y_s\middle|\bm{c}_s,\bm{c}_\ast,y_\ast\right)$ necessitates retraining the GPC-based PLA model, resulting in high computational complexity with $O(M_1n^3)$. To address these challenges, we propose ALU-based and SALU-based fingerprint selection algorithms. These algorithms aim to: 1) decrease the number of fingerprint samples required through importance sampling, and 2) eliminate the need for retraining of the GPC-based PLA model by utilizing joint distribution calculations.
\end{rem}

\begin{defn}
Motivated by the Mean objective Cost of Uncertainty (MOCU) \cite{yoon2013quantifying}, we define ALU as the expected loss difference between the OBC and the optimal classifier, formulated as
\begin{equation}
\label{22}
\begin{aligned}
& \mathcal{A}\left(\pi\left(\bm{f}\right)\right)\triangleq\mathbb{E}_{\bm{c}_s}\{1-\max_{y_s}{p\left(y_s\middle|\bm{c}_s\right)}\} \\
& -\mathbb{E}_{\pi\left(\bm{f}\right)}\left\{\mathbb{E}_{\bm{c}_s}\left[1-\max_{y_s}{p\left(y_s\middle|\bm{c}_s,\bm{f}\right)}\right]\right\}.
\end{aligned}
\end{equation}
\end{defn}
The ALU reduction is utilized as the acquisition function, given by
\begin{equation}
U^A\left(\bm{c}_\ast;\pi\left(\bm{f}\right)\right)=\mathcal{A}\left(\pi\left(\bm{f}\right)\right)-\mathbb{E}_{y_\ast|\bm{c}_s}\left[\mathcal{A}\left(\pi\left(\bm{f}|\bm{c}_\ast,y_\ast\right)\right)\right].
\end{equation}

According to \cite{zhao2021efficient}, we can derive
\begin{equation}
\label{24}
\begin{aligned}
& U\left(\bm{c}_\ast\right)=\mathbb{E}_{\bm{c}_s}\left\{1-\max_{y_s}{p\left(y_s\middle|\bm{c}_s\right)}\right\} \\
& -\mathbb{E}_{y_\ast|\bm{c}_\ast} \{\mathbb{E}_{\bm{c}_s}[1-\max_{y_s}{p\left(y_s\middle|\bm{c}_s,\bm{c}_\ast,y_\ast\right)}]\} \\
& =\mathbb{E}_{\bm{c}_s}\left\{\mathbb{E}_{y_\ast|\bm{c}_\ast}\left[\max_{y_s}{p\left(y_s\middle|\bm{c}_s,\bm{c},y\right)}\right]-\max_{y_s}{p\left(y_s\middle|\bm{c}_s\right)}\right\},
\end{aligned}
\end{equation}
which represents the expected error reduction of the OBC \cite{roy2001toward}.

\begin{defn}
Drawing inspiration from Soft-MOCU \cite{zhao2021bayesian}, we further introduce SALU in (\ref{25}), which serves as a smooth concave approximation of ALU.
\begin{equation}
\label{25}
\begin{aligned}
& \mathcal{A}^S\left(\pi\left(\bm{f}\right)\right)\triangleq\mathbb{E}_{\bm{c}_s} { \{1-\frac{1}{k} \mathtt{LogSumExp} (k\cdot p\left(y_s\middle|\bm{c}_s\right)) \} } \\
& -\mathbb{E}_{\pi\left(\bm{f}\right)} \{ \mathbb{E}_{\bm{c}_s}[1-\max_{y_s}{p\left(y_s\middle|\bm{c}_s,\bm{f}\right)}] \}.
\end{aligned}
\end{equation}
\end{defn}

The corresponding acquisition function is denoted as (\ref{softacquisitionfunction}), which has no maximization operators. After defining acquisition functions as (\ref{24}) or (\ref{softacquisitionfunction}), importance sampling and joint distribution calculation are employed as follows.

\begin{figure*}
\begin{equation}
\label{softacquisitionfunction}
U^S(\bm{c}_*)=\mathbb{E}_{\bm{c}_s}
{
\left\{
\mathbb{E}_{y_* \mid \bm{c}_*}\left[\frac{1}{k} \log \operatorname{SumExp}(k \cdot p(y_s \mid \bm{c}_s, \bm{c}_*, y_*))\right]
-
\frac{1}{k} \log \operatorname{SumExp}(k \cdot p(y_s \mid \bm{c}_s))
\right\}
}
\end{equation}
\end{figure*}

\subsubsection{Importance Sampling}
$U\left(\bm{c}_\ast\right)=\mathbb{E}_{\bm{c}_s\sim p(\bm{c}_s)}[g(\bm{c}_s;\bm{c}_*)]$ can be reformulated using a different distribution $\widetilde{p}(\bm{c}_s;\bm{c}_\ast)$ as
\begin{equation}
\label{27}
U\left(\bm{c}_\ast\right)=\mathbb{E}_{\bm{c}_s \sim \widetilde{p}\left(\bm{c}_s;\bm{c}_\ast\right)}\left[\frac{p\left(\bm{c}_s\right)g\left(\bm{c}_s;\bm{c}_\ast\right)}{\widetilde{p}\left(\bm{c}_s;\bm{c}_\ast\right)}\right]
\end{equation}
Assuming $\widetilde{p}\left(\bm{c}_s;\bm{c}_\ast\right)\propto k(\bm{c}_s;\bm{c}_\ast)p(\bm{c}_s)$, we can deduce that $p\left(y_s\middle|\bm{c}_s,\bm{c}_\ast,y_\ast\right)\approx0$ and $g\left(\bm{c}_s;\bm{c}_\ast\right)\approx0$.

Considering that $k(\bm{c}_s;\bm{c}_\ast)\approx0$, that is, $(\bm{c}_s;y_s)$ and $(\bm{c}_\ast;y_\ast)$ are independent, we can obtain $p\left(y_s\middle|\bm{c}_s,\bm{c}_\ast,y_\ast\right)\approx0$ and $g\left(\bm{c}_s;\bm{c}_\ast\right)\approx0$.

\subsubsection{Joint Distribution Calculation}
To tackle with the second challenge mentioned in \textbf{Remark \ref{remark3}}, we derive $p\left(y_s\middle|\bm{c}_s,\bm{c}_\ast,y_\ast\right)$ as
\begin{equation}
\label{jointdistributioncalculation}
p\left(y_s\middle|\bm{c}_s,\bm{c}_\ast,y_\ast\right)=\frac{p\left(y_s,y_\ast\middle|\bm{c}_s,\bm{c}_\ast\right)}{p\left(y_\ast\middle|\bm{c}_\ast\right)}.
\end{equation}
The calculation of this joint distribution requires knowledge of $p\left(y_s,y_\ast\middle|\bm{c}_s,\bm{c}_\ast\right)$.

\begin{prop}
\label{prop7}
The joint distribution $p\left(y_s,y_\ast\middle|\bm{c}_s,\bm{c}_\ast\right)$ can be expressed as
\begin{equation}
\label{29}
\begin{aligned}
& p\left(y_s=1,y_\ast=1\middle|\bm{c}_s,\bm{c}_\ast\right) \\
& =\int\Phi\left(\frac{{\widetilde{\mu}}_\ast\left(f_s\right)}{\sqrt{{\widetilde{\sigma}}_{\ast\ast}+1}}\right)\Phi\left(f_s\right)\phi\left(f_s\middle|\mu_s,\sigma_{ss}\right)df_s,
\end{aligned}
\end{equation}
where $f_s=f(\bm{c}_s)$ and $f_\ast=f(\bm{c}_\ast)$. The variables $\mu_{s^\ast}$ and $\Sigma_{s^\ast}$ denote the marginal mean vector and covariance matrics of $f_s$ and $f_\ast$, respectively. Let $\mu_{s^\ast}=(\begin{matrix}\mu_s\\\mu_\ast\\\end{matrix})$ and $\Sigma_{s^\ast}=(\begin{matrix}\sigma_{ss}&\sigma_{s^\ast}\\\sigma_{s^\ast}&\sigma_{\ast\ast}\\\end{matrix})$, we have $\phi\left(f_s,f_\ast\middle|\mu_{s^\ast},\Sigma_{s^\ast}\right)=\left(f_s\middle|\mu_s,\sigma_{ss}\right)\left(f_\ast\middle|{\widetilde{\mu}}_\ast(f_s),{\widetilde{\sigma}}_{\ast\ast}\right)$, where ${\widetilde{\mu}}_\ast\left(f_s\right)=\mu_\ast+\frac{(f_s-\mu_s)\sigma_{s^\ast}}{\sigma_{ss}}$ and ${\widetilde{\sigma}}_{\ast\ast}=\sigma_{\ast\ast}-\frac{\sigma_{s^\ast}^2}{\sigma_{ss}}$.
\end{prop}

\begin{IEEEproof}
    See Appendix \ref{appf}.
\end{IEEEproof}

The detailed processes of the designed (S)ALU-based fingerprint selection algorithms are depicted in \textbf{Algorithm 3}.

\begin{algorithm}[!t]
\caption{Process of the designed ALU-based and SALU-based Fingerprint Selection Algorithms}
\textbf{Input:}  $p(\bm{c})$; $q(\bm{f}|C,\bm{y})$

\textbf{Process:}
\begin{algorithmic}[1]
\label{algorithm3}
\State Sampling $M_1$ fingerprints of $\bm{c}_\ast \sim p(\bm{c})$
\For{each $\bm{c}_\ast$}
\State Calculating $p\left(y_\ast\middle|\bm{c}_\ast\right)$ by (\ref{20})
\State Sampling $M_2$ fingerprints of $\bm{c}_s \sim \widetilde{p}(\bm{c}_s;\bm{c}_\ast)$
\For{$y_\ast$ in $\{0,1\}$}
\For{each $\bm{c}_s$}
\State Calculating $p\left(y_s\middle|\bm{c}_s,\bm{c}_\ast,y_\ast\right)$ by (\ref{29})
\State Calculating $p\left(y_s\middle|\bm{c}_s\right)$ by (\ref{20})
\State Calculating $p\left(y_s\middle|\bm{c}_s,\bm{c}_\ast,y_\ast\right)=\frac{p\left(y_s,y_\ast\middle|\bm{c}_s,\bm{c}_\ast\right)}{p\left(y_s\middle|\bm{c}_s\right)}$
\State Calculating $g(\bm{c}_s;\bm{c}_\ast)$
\EndFor
\EndFor
\State $U\left(\bm{c}_\ast\right)=\frac{1}{M_2}\sum_{\bm{c}_s}\frac{p(\bm{c}_s)g(\bm{c}_s;\bm{c}_\ast)}{\widetilde{p}(\bm{c}_s;\bm{c}_\ast)}$
\EndFor

\textbf{Output:} $\widetilde{\bm{c}}=\arg \max_{\bm{c}_\ast}U\left(\bm{c}_\ast\right)$
\end{algorithmic}
\end{algorithm}

\section{Performance Evaluation and Simulation Results}
\label{simulation}
\subsection{Performance Metric}
The authentication performance is evaluated using the authentication error rate $R_e$, which is calculated as follows:
\begin{equation}
    R_e=\frac{1}{N_a}\sum_{n=1}^{N_a}\mathbb{I}\left(\bm{L}_n \neq \bm{Y}_n\right),
\end{equation}
where $N_a$ denotes the total number of testing fingerprints, $\bm{L}_n$ and $\bm{Y}_n$ respectively represent the true and estimated identities of the $n$th fingerprint, and $\mathbb{I}(\cdot)$ is an indicator function that returns 1 if $\cdot$ is true and 0 otherwise.

\begin{lem}
\label{lem}
$R_e$ is a fine-grained metric that incorporates both miss detection rate $P_{md}$ and false alarm rate $P_{fa}$. $R_e$ can be further denoted as
\begin{equation}
R_e=
\displaystyle\frac
{1}
{1+
{\displaystyle\frac{{TL}/{TA}+1}{{\displaystyle\frac{1}{\displaystyle\frac{1}{P_{md}}-1}+\frac{1}{\displaystyle\frac{1}{P_{fa}}-1}}}}},
\end{equation}
where $TL$ and $TA$ are illustrated in Tab. \ref{table2} \cite{meng2023physical,xia2021multiple}.
\end{lem}
\begin{IEEEproof}
    See Appendix \ref{appg}.
\end{IEEEproof}

\begin{table}[htb] 
\begin{center}   
\caption{Confusion Matrix of Alice and Eve}
\label{table2}
\setlength{\tabcolsep}{4pt}
\begin{tabular}{|c|c|c|}
\hline
\textbf{Actual\textbackslash{}Predicted} & \textbf{Alice} & \textbf{Eve} \\ \hline
\textbf{Alice} & \makecell{ True Legitimate (TL) } & \makecell{False Attack (FA)} \\ \hline
\textbf{Eve} & \makecell{False Legitimate (FL)} & \makecell{True Attack (TA)} \\ \hline
\end{tabular}
\end{center}
\end{table}

\subsection{Simulation Parameters}

\begin{figure}[t]
\centering
\includegraphics[width=0.3\textwidth]{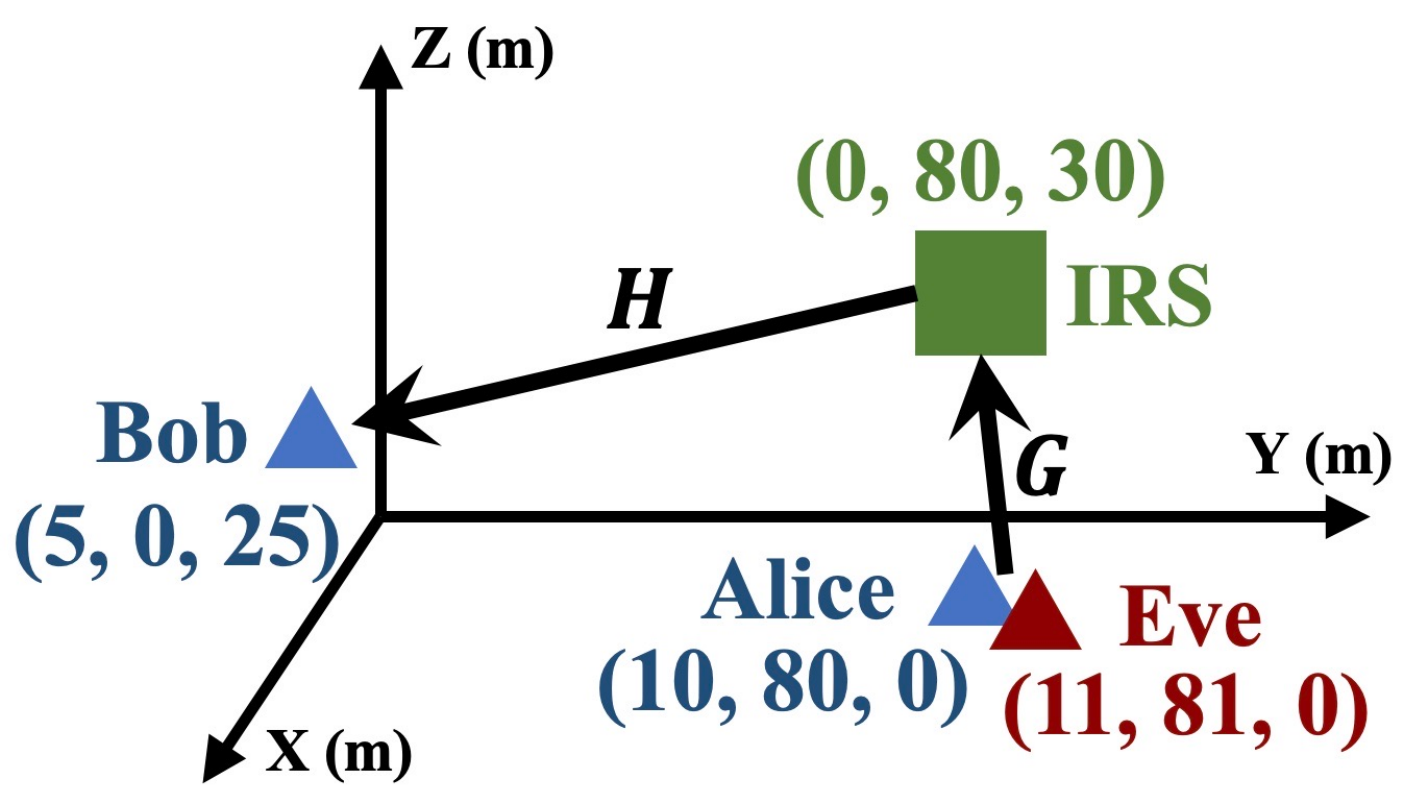}
\caption{The three-dimensional spatial positions of involved nodes, including Alice, RIS, Bob, and Eve.}
\label{fig3}
\vspace{-1em}
\end{figure}
Configurable fingerprint samples are generated using MATLAB platform. The three-dimensional space positions of involved nodes are depicted in Fig. \ref{fig3}. Alice, Eve, and Bob are equipped with Uniform Linear Arrays (ULAs), with Alice/Eve having $N_T=2$ arrays and Bob having $N_R=4$ arrays. The RIS consists of a Uniform Rectangular Array (URA) composed of $N=N_y N_z=8*32$ reflective elements, with each element spaced half a wavelength apart. The channels $\bm{Q}^{(R,B)}$ and $\bm{Q}^{(A,R)}$ are modeled as Rician channels, given by
\begin{equation}
\bm{Q}^{(R,B)}=\sqrt{\frac{PL \kappa_{RB}}{1+\kappa_{RB}}} \overline{\bm{Q}^{(R,B)}}+\sqrt{\frac{PL}{1+\kappa_{RB}}} \widetilde{\bm{Q}^{(R,B)}}
\end{equation}
\begin{equation}
\bm{Q}^{(A,R)}=\sqrt{\frac{PL\kappa_{AR}}{1+\kappa_{AR}}} \overline{\bm{Q}^{(A,R)}}+\sqrt{\frac{PL}{1+\kappa_{AR}}} \widetilde{\bm{Q}^{(A,R)}}
\end{equation}
where $\overline{\bm{Q}^{(R,B)}}$ and $\overline{\bm{Q}^{(A,R)}}$ represent the Line of Sight (LoS) paths, $\widetilde{\bm{Q}^{(R,B)}}$ and $\widetilde{\bm{Q}^{(A,R)}}$ represent the Non-LoS (NLoS) paths, and $\kappa_{RB}$ and $\kappa_{AR}$ denote Rician factors. The corresponding path loss are modeled according to the 3GPP TR 38.901 \cite{3gpp2022study}.
The LoS paths are modeled according to \cite{hu2021reconfigurable}, while the NLoS paths are modeled as Rayleigh fading channels, where each element of the channel matrix follows an independent identically distributed complex Gaussian distribution $\mathcal{CN}(0,\bm{I})$.
Artificial noises $\bm{W} \sim \mathcal{CN}(0,{\sigma}_w^2\bm{I})$ are used to represent noisy scenarios, with ${\sigma}_w^2=10^{-20}$. The performance of the PLA models is evaluated using the PyCharm platform. The synthetic fingerprint datasets are divided into training and testing sets. Initially, 2 fingerprints respectively from Alice and Eve are selected for labeling and used to estimate the GPC hyperparameters. The learning process is repeated 100 times. The simulation and hyperparameters are detailed in Tab. \ref{simulationparameter}. The computer configurations include an Intel Core i5-13600KF processor with a 3.50 GHz basic frequency and 32 GB of RAM.

\begin{table}[t]
\caption {Simulation and Hyper Parameters}
\label{simulationparameter}
\begin{center}
\begin{tabular}{|c| c|} 
\hline
\textbf{Parameters}                    & \textbf{Values} \\ \hline
Bandwidth                       & 1 MHz              \\ \hline
Carrier Frequency $f_c$                & 3.5 GHz             \\ \hline
Rice factors $\kappa_{h}$ and $\kappa_{g}$                            & 3 and 4               \\ \hline
Number of RIS elements                 & 8*32            \\ \hline
Number of antennas of Bob $N_R$              & 4               \\ \hline

Number of antennas of each transmitter $N_T$ & 2               \\ \hline

Number of each transmitter's training fingerprints & 800 \\ \hline
Number of each transmitter's testing fingerprints & 200 \\ \hline
Learning rate & 0.001 \\ \hline
Iteration number & 30 \\ \hline
The variance of the RBF kernel & 1 \\ \hline
The length scale of the RBF kernel & 0.4 \\ \hline
\end{tabular}
 \end{center}
\end{table}

\subsection{Baseline Algorithms}
The baseline fingerprint selection algorithms for performance comparison are as follows.
\begin{itemize}
    \item Random Selection (RS): The RS algorithm employs a straightforward approach by choosing fingerprint samples for labeling in a completely random manner. In this process, every sample within the dataset is assigned an equal probability of being selected, ensuring an unbiased and uniform sampling distribution. This method does not prioritize any specific characteristics or patterns in the data, relying solely on chance to determine which samples are labeled.
    \item Maximum Entropy Sampling (MES)\cite{sebastiani2000maximum}: The MES algorithm is based on information theory, specifically Shannon entropy, which quantifies the uncertainty or information content in a probability distribution. In the context of active learning, MES aims to select a subset of samples that maximizes the entropy of the selected set, thereby minimizing uncertainty and maximizing information gain.
    \item Bayesian Active Learning by Disagreement (BALD) \cite{houlsby2011bayesian}: The BALD algorithm operates by maximizing the disagreement among posterior parameters with respect to their predicted outcomes. This disagreement refers to the variability or uncertainty in the model's predictions, which stems from the uncertainty in the posterior distribution of the model parameters. Specifically, it quantifies the epistemic uncertainty—the uncertainty arising from the model's incomplete knowledge about the true parameters given the observed data. By targeting samples where this disagreement is highest, BALD identifies the most informative data points for labeling, thereby efficiently reducing uncertainty and improving model performance.
\end{itemize}

\subsection{Simulation Results}

\begin{figure}[t]
\centering
\includegraphics[width=0.4\textwidth]{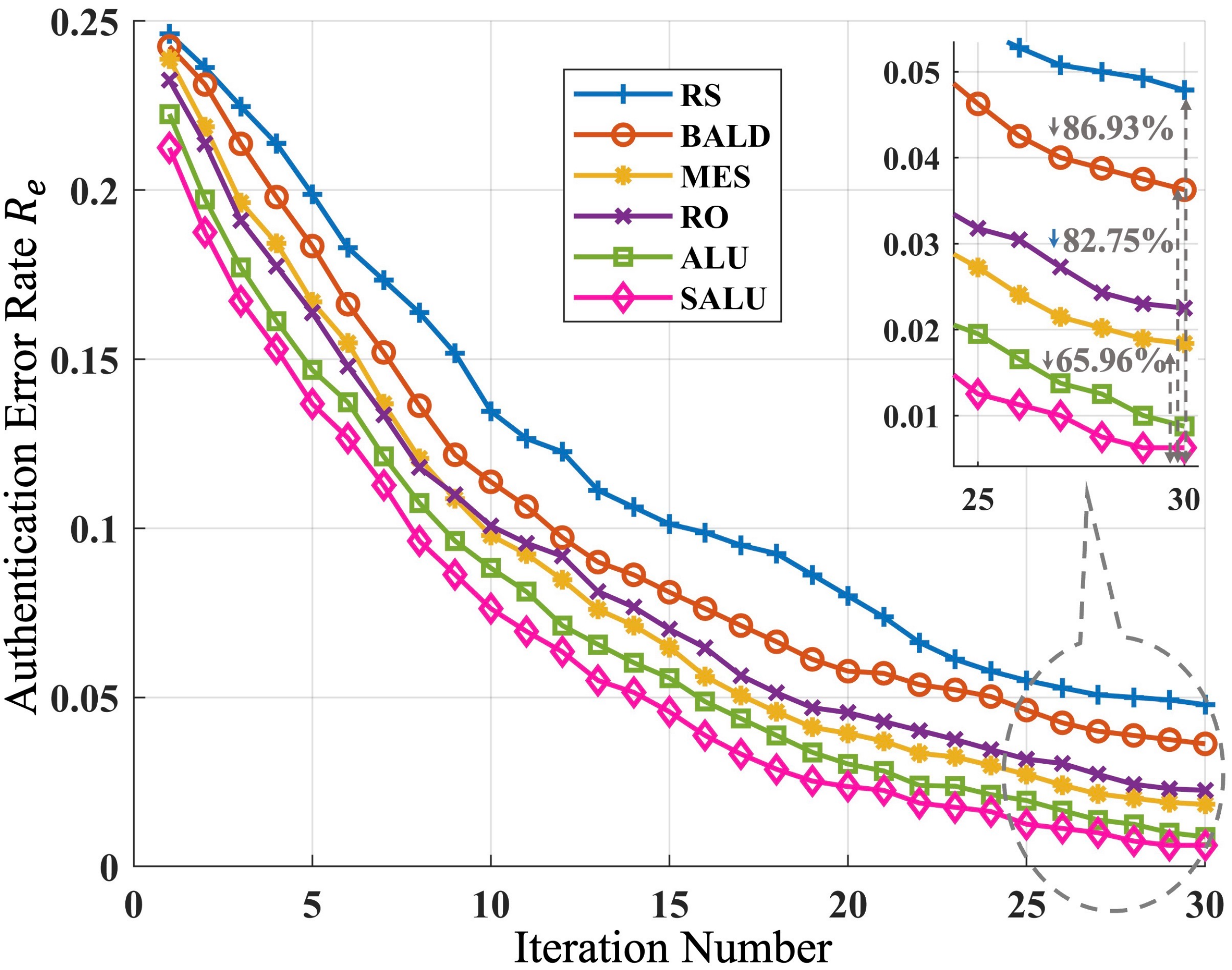}
\caption{Illustration of the authentication error rate $R_e$ compared across various iteration numbers in RIS-assisted wireless environments.}
\label{fig4}
\vspace{-1em}
\end{figure}

\textbf{Comparisons with baseline algorithms.} Fig. \ref{fig4} compares the proposed fingerprint selection algorithms with three baseline active learning algorithms. As the iteration number increases, the proposed GPC-based PLA model observes more labeled fingerprint samples, gains a deeper understanding of configurable fingerprint distributions, and achieves more accurate predictions, thereby reducing authentication error rates. The analysis of the fingerprint selection algorithms is as follows. The RS algorithm selects fingerprints randomly, potentially choosing samples with similar characteristics repeatedly, limiting the PLA model's ability to learn the fingerprint distribution effectively. The BALD algorithm excels in learning the characteristics of low-dimensional data rather than high-dimensional fingerprints. The MES algorithm tends to prioritize querying points near decision boundaries, but multi-dimensional cascaded fingerprints have multiple boundaries. The proposed ALU algorithm uses expected error reduction as the acquisition function, outperforming the proposed RO algorithm in authentication performance. Compared to the ALU algorithm, the suggested SALU algorithm achieves a lower error rate due to its smoother concave approximation. Additionally, compared with the three baseline algorithms, the SALU algorithm reduces authentication error rates by 65.96\% to 86.93\%. Fig. \ref{fig4} demonstrates the superior sampling efficiency of the proposed fingerprint selection algorithms over baseline active learning methods.

\begin{figure}[t]
\centering
\includegraphics[width=0.4\textwidth]{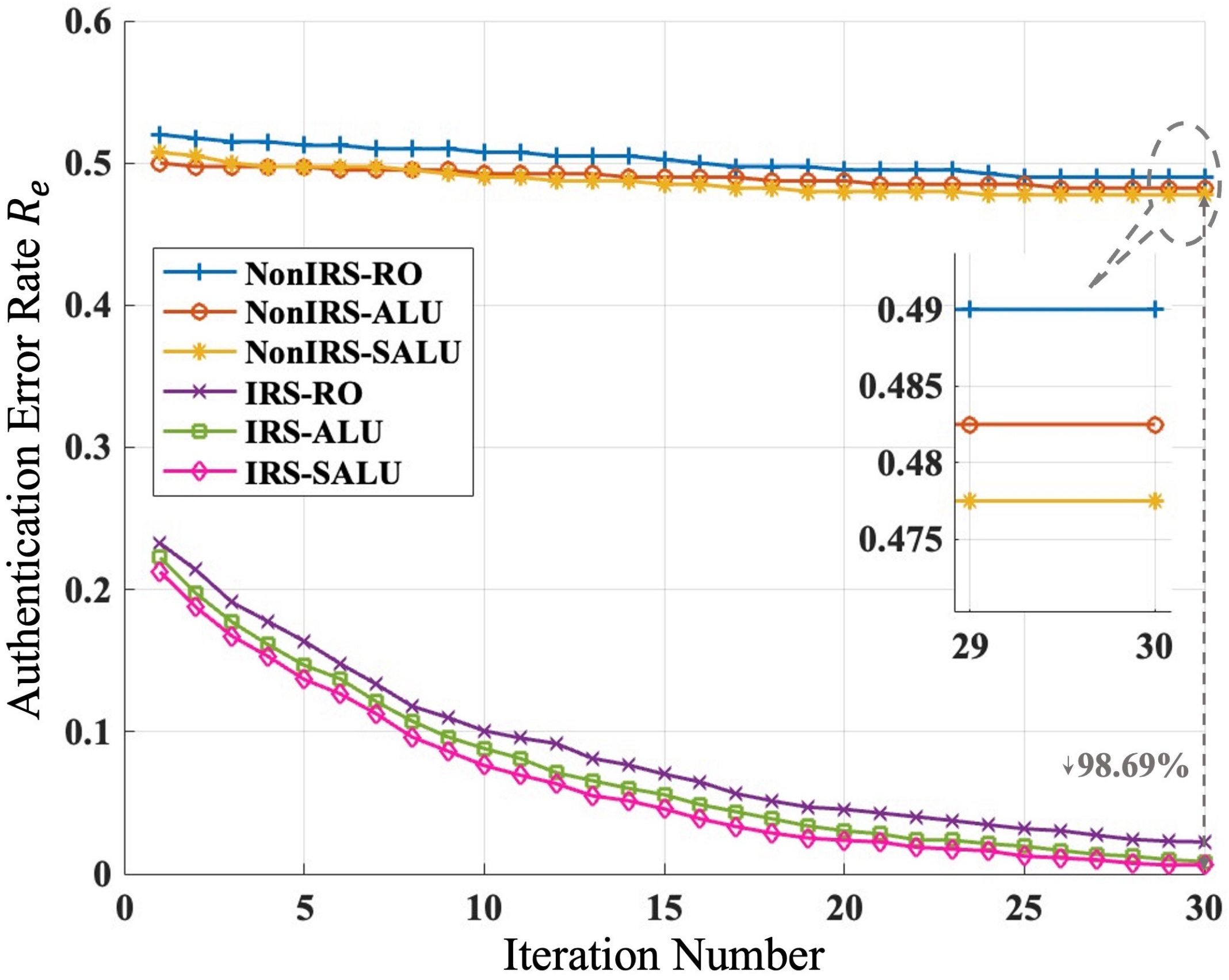}
\caption{Comparison of $R_e$ across virous iteration numbers in both non-RIS and RIS-aided scenarios.}
\label{fig5}
\vspace{-1em}
\end{figure}

\textbf{Effectiveness of RISs.} Fig. \ref{fig5} presents the authentication performance in non-RIS and RIS-assisted wireless environments. In non-RIS environments with low SNRs, the channels between Alice/Eve and Bob are modeled as Rayleigh channels, characterized by severe fading. Across all iterations, the authentication error rate of the proposed fingerprint selection algorithms remains consistently around 0.5 in non-RIS scenarios. Due to the severe fading of the Rayleigh channel, there is significant overlap in the channel fingerprint space of Alice and Eve. Despite the gradual increase in labeled fingerprint samples, effectively learning the inherent characteristics of fingerprints remains challenging for the PLA model. In contrast, the channel quality improves with the transmission paths created by RISs. With increasing iterations, an adequate number of labeled fingerprint samples significantly enhance authentication performance. Therefore, leveraging RIS assistance, the PLA model achieves superior authentication performance. For instance, compared to scenarios without RIS, the authentication error rate of the SALU algorithm with RIS assistance decreases by 98.69\%. Fig. \ref{fig5} shows the effectiveness of RISs in enhancing authentication performance compared with the conventional non-RIS-based PLA framework under low SNR environments.

\begin{figure}[t]
\centering
\includegraphics[width=0.4\textwidth]{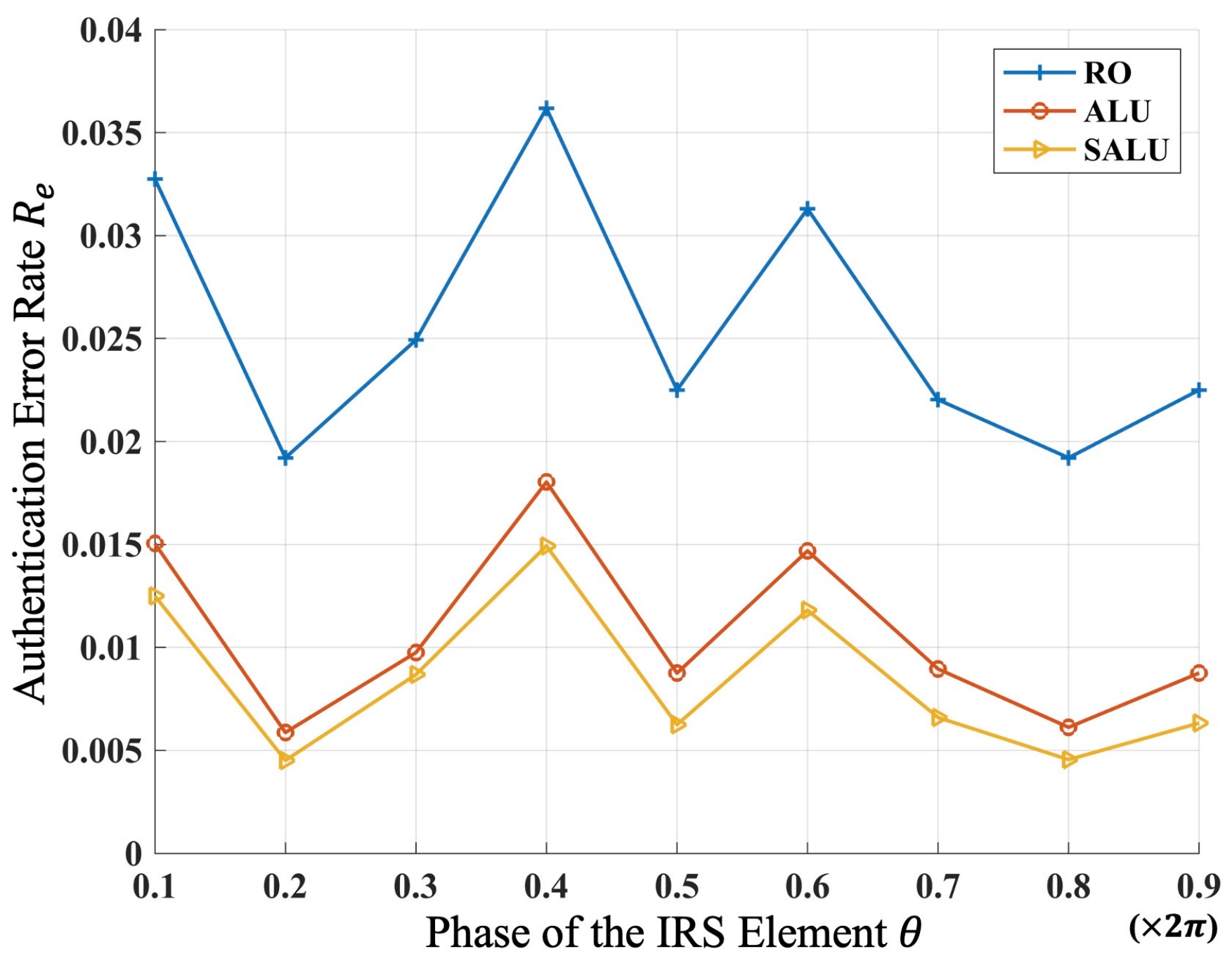}
\caption{Illustration of $R_e$ compared across various of RIS phase parameters $\theta$.}
\label{fig6}
\vspace{-1em}
\end{figure}

\textbf{Performance versus RIS phase parameters.} Fig. \ref{fig6} illustrates the authentication performance across various phase parameters of the RIS element, denoted as $\theta=\theta_1=\theta_2=...=\theta_n$, where $\theta_n$ represents the phase parameter of the $n$th RIS element. As the phase increases, the authentication error rate displays fluctuations in amplitude. However, drawing definitive conclusions  about the relationship between RIS phase parameters and authentication error rates from Fig. \ref{fig6} is challenging. The optimization of RIS parameters remains an important area of research. Fig. \ref{fig6} confirms that the choice of RIS phase parameters has a significant impact on authentication performance.

\begin{figure}[t]
\centering
\includegraphics[width=0.4\textwidth]{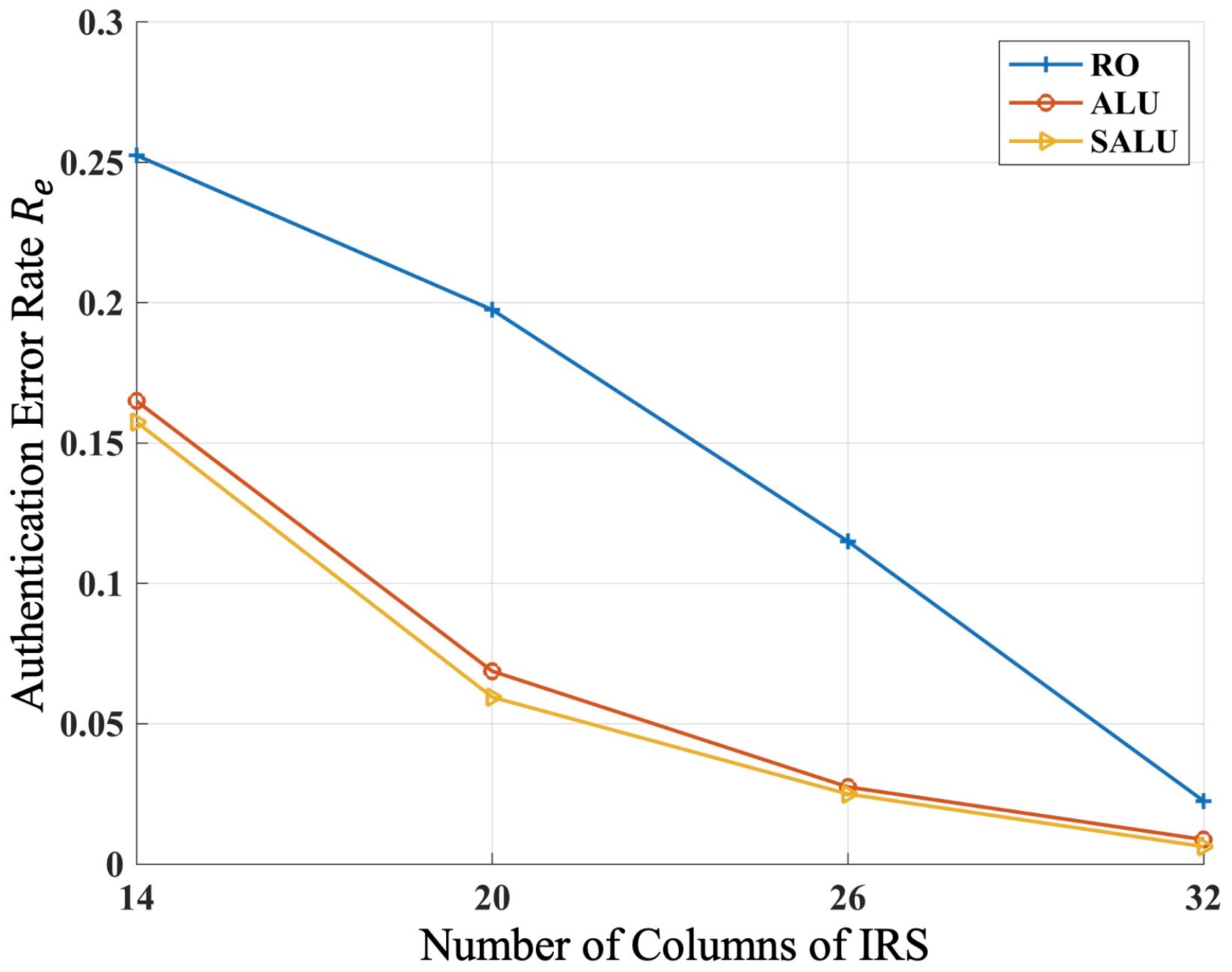}
\caption{Illustration of $R_e$ compared across various RIS column numbers.}
\label{fig7}
\vspace{-1em}
\end{figure}

\textbf{Performance versus RIS columns numbers.}
Fig. \ref{fig7} compares the authentication performance across different numbers of RIS columns, denoted as $N_z$. As $N_z$ increases, the total number of RIS element, $N=N_yN_z$, also increases. Consequently, the dimensions of $\bm{H}$, $\bm{\Psi}$, and $\bm{G}$ within the cascade channel matrix $\bm{Q}^{(A,I,B)}=\bm{H \Psi G}$ expand. When the dimensions of $\bm{H}$, $\bm{\Psi}$, and $\bm{G}$ are small, the limited variability in fingerprint distribution may not suffice for consistent transmitter differentiation. Conversely, with larger dimensions, it becomes more challenging for attackers to predict or imitate each component of the fingerprints. Fig. \ref{fig7} confirms that a greater number of RIS columns contributes to enhanced authentication performance.

\begin{figure}[t]
\centering
\includegraphics[width=0.4\textwidth]{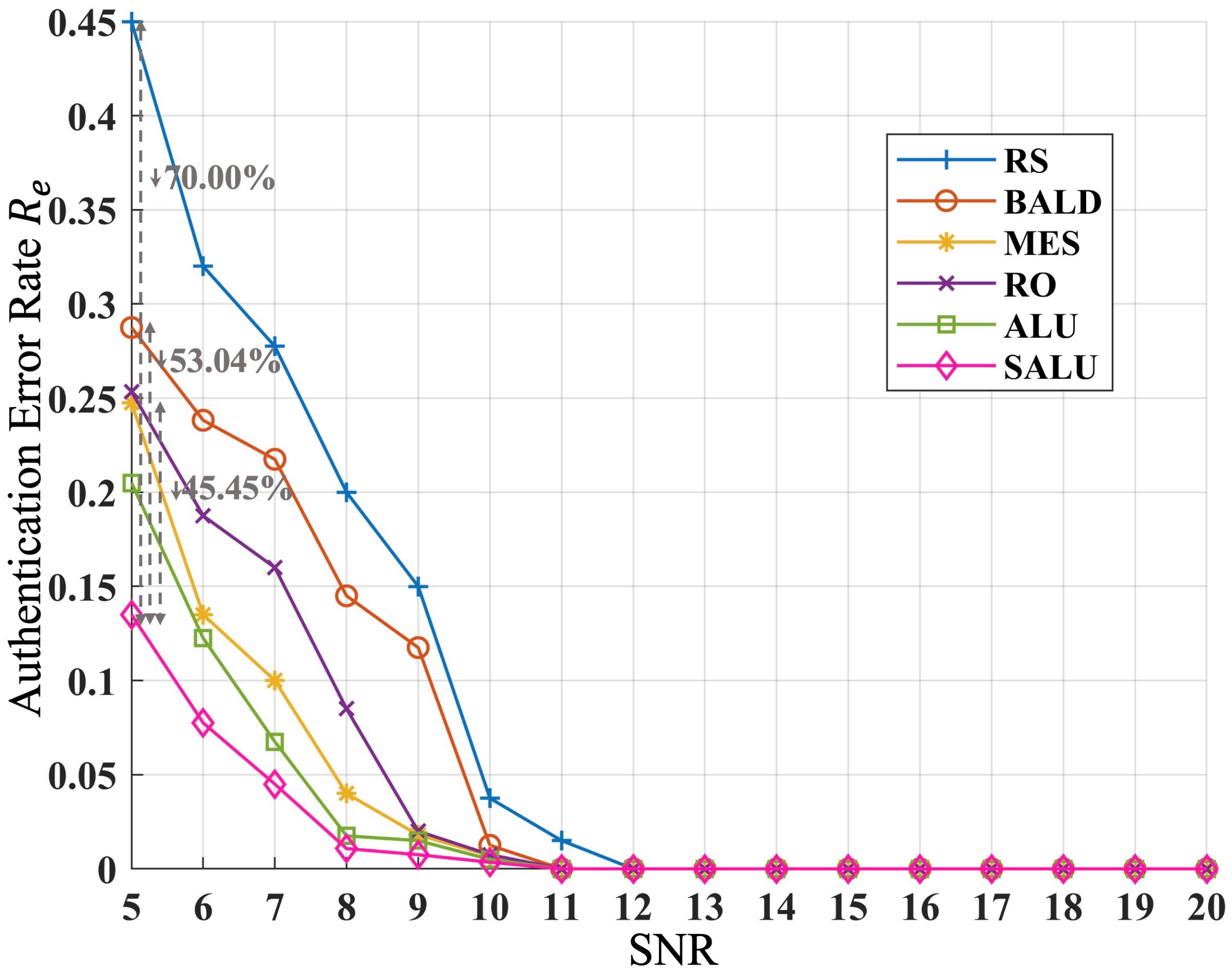}
\caption{Illustration of $R_e$ compared across various SNRs.}
\label{fig8}
\vspace{-1em}
\end{figure}

\textbf{Performance under different SNRs.}
Although Fig. \ref{fig5} demonstrates the efficacy of RISs in improving authentication performance, channel fingerprint estimation is susceptible to biases from noises in channels between Alice/Eve and RISs, or between RISs and Bob. To analyze authentication performance under varying channel estimation errors, artificial noises of different powers are considered, as depicted in Fig. \ref{fig8}.
For $\text{SNR}\ge12$, the distinction between Alice’s and Eve’s fingerprints is pronounced, enabling accurate transmitter differentiation with all algorithms.
At $\text{SNR}=11$, only the RS algorithm can not accurately identify all fingerprints due to its limited fingerprint selection capability.
As $\text{SNR}\le10$, decreasing SNR causes Alice’s and Eve’s fingerprint spaces to increasingly overlap, posing challenges for the PLA model in transmitter identification. Nevertheless, compared to baseline algorithms, the proposed SALU algorithm reduces authentication error rate by 45.45\% to 70.00\% at $\text{SNR}=5$. Fig. \ref{fig8} verifies the robustness of the proposed algorithms against channel estimation errors.

\begin{figure}[t]
\centering
\includegraphics[width=0.4\textwidth]{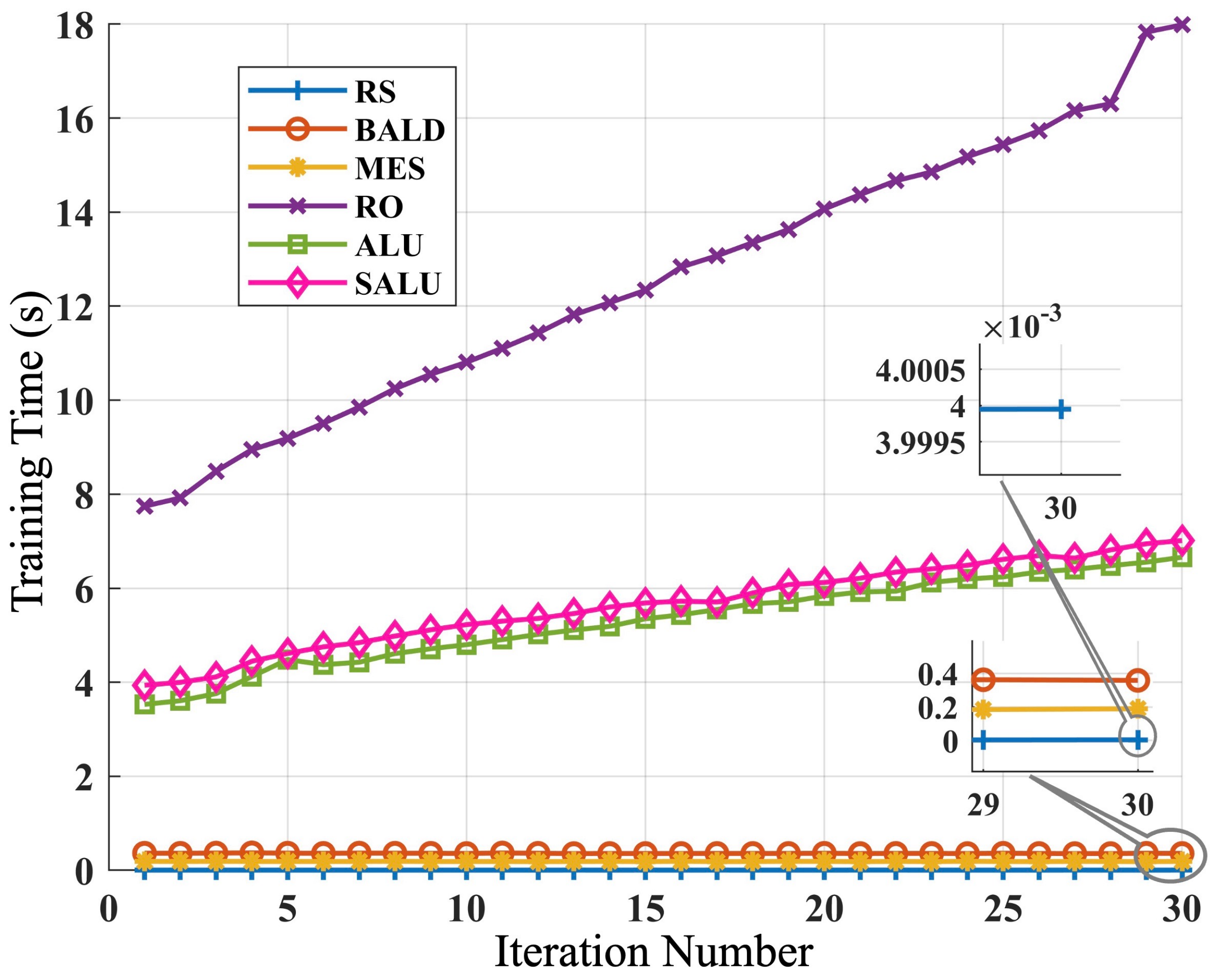}
\caption{Illustration of the training time compared across various iteration numbers under the RIS-assisted wireless environment.}
\label{fig9}
\vspace{-1em}
\end{figure}

\textbf{Training complexity.}
Fig. \ref{fig9} employs computational time as a metric to illustrate the computational complexity of various fingerprint selection algorithms.
The RS algorithm exhibits the lowest training time, as it selects fingerprint samples in a purely random manner, requiring no additional computational overhead.
The BALD algorithm, which quantifies sample disagreement by calculating the mutual information of model predictions, is more computationally intensive than the MES algorithm. This is because MES only requires computing the entropy of model predictions to evaluate samples, a simpler and less resource-demanding process.
The (S)ALU algorithms, which assess sample importance by calculating the expected cost of uncertainty, demonstrate higher computational complexities compared to BALD. This is due to the additional steps involved in evaluating the impact of uncertainty on the task-specific objectives.
The RO algorithm exhibits the highest computational complexity among all methods. This is attributed to the necessity of retraining the GPC-based PLA model for each new candidate pair $(\bm{c}_*,y_*)$. In contrast, the (S)ALU algorithms mitigate this computational burden by incorporating importance sampling, as outlined in (\ref{27}), and joint distribution calculations, as described in (\ref{jointdistributioncalculation}), thereby significantly accelerating the training process.

\section{Conclusion and Future Work}
\label{conclusion}
To bolster  wireless identity security in IoT systems enabled by 6G technology, we propose an EGPC-based PLA scheme that innovates over state-of-the-art approaches in several key aspects. Firstly, we employ RISs to enhance fingerprint accuracy in low SNR environments. Secondly, GPs are utilized to model configurable fingerprints, and their posterior distributions are derived for further analysis. Thirdly, we integrate active learning to achieve lightweight cross-layer authentication. We also introduce three fingerprint selection algorithms aimed at improving efficiency.
Simulations validate the effectiveness of RISs and demonstrate the superiority of our proposed algorithms. The future work is provided as follows.

\textbf{Real-World Testing:} This paper employs MATLAB to generate synthetic configurable fingerprints under ideal conditions, to which artificial noise is added to simulate fingerprints under varying SNR conditions. However, compared to real-world fingerprints, synthetic fingerprints may fail to account for certain complex factors in the actual environment, such as nonlinear effects. Moreover, synthetic fingerprints are often static or derived from historical data, making it challenging to capture real-time channel variations, which results in deviations from real-world fingerprints. In the future, real-world configurable fingerprints obtained through techniques such as compressed sensing, matrix factorization, or deep learning will be utilized to validate the effectiveness of PLA models.

\textbf{Deployment of RISs in IoT:} The integration of RISs into IoT can enhance the reliability of channel fingerprints, thereby boosting authentication performance, while also improving the energy efficiency of communication systems. However, deploying RISs may introduce additional signal interference, particularly in densely populated IoT environments, rendering interference management a pressing challenge. In the future, such interference can be effectively mitigated through the joint optimization of RIS and IoT device transmission strategies, coupled with the adoption of advanced interference management techniques like beamforming and resource allocation. 
Furthermore, the integration of RISs may also introduce new security threats. For instance, attackers could potentially gain control over RISs and maliciously adjust RIS parameters to manipulate configurable channel fingerprints, thereby enabling illegal fingerprints to be authenticated as legitimate. This highlights the need for robust security mechanisms to protect RIS configurations from unauthorized access and tampering. Additionally, the privacy implications of configurable fingerprints must be carefully studied to prevent exploitation by adversaries.

\textbf{Optimization of RIS Parameters:} It is noteworthy that optimizing RIS parameters remains a topic requiring further exploration, which we intend to investigate using deep RL algorithms in the future. To begin with, the optimization problem associated with RIS parameters is formulated as a deep RL framework. This framework encompasses several key components: the state space, which includes variables such as configurable fingerprints, user locations, transmit power levels, and SNRs; the action space, which consists of the adjustable parameters of RISs and detection thresholds; the reward function, which is designed to quantify authentication performance; and the environment, which represents the dynamic wireless communication scenarios. Following this, a state encoder is developed to transform high-dimensional configurable fingerprints into compact, low-dimensional feature representations, facilitating efficient processing. A policy network is then constructed to generate optimal RIS parameter configurations based on the current state. Additionally, a value network is implemented to assess the long-term utility or value of the current state, aiding in the decision-making process. Finally, the system undergoes policy updates and model training using the real-world dataset.

\textbf{Scalability of the Proposed Scheme:} In the context of large-scale IoT networks, the PLA problem can still be formulated as a multi-class classification task. However, the computational complexity associated with calculating the expected error reduction becomes significantly higher due to the increased number of IoT devices, leading to substantial computational overhead.
To mitigate this challenge, the proposed EGPC-based PLA framework can be effectively scaled to incorporate uncertainty sampling strategies, thereby reducing computational demands. Specifically, the framework leverages the following uncertainty sampling techniques: Least Confidence Sampling: Selecting the sample for which the model has the lowest prediction probability, indicating high uncertainty in the model's decision. Margin Sampling: Choosing samples where the prediction probabilities of the top two classes are the closest, reflecting ambiguity in classification. Entropy Sampling: Identifying the sample with the largest entropy in the predicted probability distribution, which represents the highest overall uncertainty.
By adopting these uncertainty sampling methods, the EGPC-based PLA framework not only maintains its effectiveness in large-scale IoT networks but also significantly reduces the computational burden, making it a practical and scalable solution for real-world deployment.

\textbf{Generalization for Different IoT Scenarios:} 
This paper leverages channel fingerprints as identity features to implement access authentication. In the future, these identity features can be substituted with other physical-layer quantities to better suit diverse IoT scenarios. For instance, body movement characteristics could be utilized in body area networks, while multidimensional perceptual features might be employed in unmanned aerial vehicle (UAV) applications. Such adaptability ensures the proposed EGPC-based framework remains versatile and effective across a wide range of IoT environments.

\appendices
\section{The Proof of Proposition \ref{prop2}}
\label{appa}
The jointly Gaussian random vectors $\bm{c}$ and $\bm{y}$ satisfy
\begin{equation}
\begin{aligned}
&    \left[\begin{matrix}\bm{c}\\\bm{y}\\\end{matrix}\right] \sim \mathcal{N}\left(\left[\begin{matrix}\bm{\mu}_c\\\bm{\mu}_y\\\end{matrix}\right],\left[\begin{matrix}A&D\\D^T&B\\\end{matrix}\right]\right) \\
& = \mathcal{N}\left(\left[\begin{matrix}\bm{\mu}_c\\\bm{\mu}_y\\\end{matrix}\right],\left[\begin{matrix}\widetilde{A}&\widetilde{D}\\{\widetilde{D}}^T&B\\\end{matrix}\right]^{-1}\right).
\end{aligned}
\end{equation}
According to \cite{williams2006gaussian}, the marginal and conditional distribution of $\bm{c}$ are shown as
\begin{equation}
    \bm{c} \sim \mathcal{N}\left(\bm{\mu}_c,A\right)
\end{equation}
and
\begin{equation}
\label{app2-1}
    \bm{c}|\bm{y} \sim \mathcal{N}\left(\bm{\mu}_c+DB^{-1}\left(\bm{y}-\bm{\mu}_y\right),A-DB^{-1}D^T\right)
\end{equation}
or
\begin{equation}
\label{app2-2}
    \bm{c}|\bm{y} \sim \mathcal{N}\left(\bm{\mu}_c-{\widetilde{A}}^{-1}\widetilde{D}\left(\bm{y}-\bm{\mu}_y\right),{\widetilde{A}}^{-1}\right).
\end{equation}

Thus, \textbf{Proposition \ref{prop2}} is proved.

\section{The Proof of Proposition \ref{prop3}}
\label{appb}
The product of two Gaussian distributions can be given by
\begin{equation}
\mathcal{N}\left(\bm{c}\middle|\bm{a},A\right)\mathcal{N}\left(\bm{c}\middle|\bm{b},B\right)=Z^{-1}\mathcal{N}\left(\bm{c}\middle|\bm{c},D\right),
\end{equation}
where
\begin{equation}
\label{app4}
    \bm{c}=D\left(A^{-1}\bm{a}+B^{-1}\bm{b}\right),
\end{equation}
\begin{equation}
\label{app5}
    D=\left(A^{-1}+B^{-1}\right)^{-1},
\end{equation}
and
\begin{equation}
\label{app6}
    Z^{-1}=\left(2\pi\right)^{-\frac{D}{2}}\left|A+B\right|^{-\frac{1}{2}}\exp{\left(-\frac{\left(\bm{a}-\bm{b}\right)^T\left(\bm{a}-\bm{b}\right)}{2\left(A+B\right)}\right)}.
\end{equation}

Therefore, by multiplying the cavity distribution by $t_i$ from (\ref{11}), \textbf{Proposition \ref{prop3}} is proved.

\section{The Proof of Proposition \ref{prop4}}
\label{appc}
Consider the following integral:
\begin{equation}
\label{app7}
Z=\int_{-\infty}^{\infty}{\Phi\left(\frac{c-m}{v}\right)\mathcal{N}(c|\mu,\sigma^2)dc},
\end{equation}
where
\begin{equation}
    \Phi\left(c\right)=\int_{-\infty}^{c}{\mathcal{N}\left(y\right)dy}.
\end{equation}

For $v>0$, by substituting $ z=y-c+\mu-m$ and $w=c-\mu$, we obtain:
\begin{equation}
\begin{aligned}
& Z_{v>0}=\frac{\int_{-\infty}^{\infty}\int_{-\infty}^{c}\exp{\left(-\frac{\left(y-m\right)^2}{2v^2}-\frac{\left(c-\mu\right)^2}{2\sigma^2}\right)}}{2\pi\sigma v}dydc \\
& =\frac{\int_{-\infty}^{\mu-m}\int_{-\infty}^{\infty}\exp{\left(-\frac{\left(z+w\right)^2}{2v^2}-\frac{w^2}{2\sigma^2}\right)}}{2\pi\sigma v}dwdz
\end{aligned}
\end{equation}
and
\begin{equation}
\begin{aligned}
& Z_{v>0} \\
& =\frac{\int_{-\infty}^{\mu-m}\int_{-\infty}^{\infty}\exp{\left(-\frac{1}{2}\left[\begin{matrix}w\\z\\\end{matrix}\right]^T\left[\begin{matrix}\frac{1}{v^2}+\frac{1}{\sigma^2}&\frac{1}{v^2}\\\frac{1}{v^2}&\frac{1}{v^2}\\\end{matrix}\right]\left[\begin{matrix}w\\z\\\end{matrix}\right]\right)}}{2\pi\sigma v}dwdz \\
& =\int_{-\infty}^{\mu-m}\int_{-\infty}^{\infty}\mathcal{N}\left(\left[\begin{matrix}w\\z\\\end{matrix}\right]|\mathbf{0},\left[\begin{matrix}\sigma^2&-\sigma^2\\-\sigma^2&v^2+\sigma^2\\\end{matrix}\right]\right)dwdz.
\end{aligned}
\end{equation}
According to (\ref{app2-1}) and (\ref{app2-2}), we derive
\begin{equation}
\label{app11}
    Z_{v>0}=\frac{\int_{-\infty}^{\mu-m}\exp{\left(-\frac{z^2}{2\left(v^2+\sigma^2\right)}\right)}dz}{\sqrt{2\pi(v^2+\sigma^2)}}=\Phi\left(\frac{\mu-m}{\sqrt{v^2+\sigma^2}}\right).
\end{equation}
For $v<0$, using $\Phi\left(-z\right)=1-\Phi\left(z\right)$ and (\ref{app7}), we obtain
\begin{equation}
\label{app12}
Z_{v<0}=1-\Phi\left(\frac{\mu-m}{\sqrt{v^2+\sigma^2}}\right)=\Phi\left(-\frac{\mu-m}{\sqrt{v^2+\sigma^2}}\right).
\end{equation}

Combining (\ref{app11}) and (\ref{app12}), we get
\begin{equation}
\label{app13}
Z=\int\Phi\left(\frac{c-m}{v}\right)\mathcal{N}\left(c\middle|\mu,\sigma^2\right)dc=\Phi\left(z\right),
\end{equation}
where $z=\frac{\mu-m}{v\sqrt{1+\sigma^2/v^2}} (v\neq0)$. 
Our goal is to find the moments of
\begin{equation}
q\left(c\right)=Z^{-1}\Phi\left(\frac{c-m}{v}\right)\mathcal{N}\left(c\middle|\mu,\sigma^2\right).
\end{equation}
By differentiating with respect to $\mu$ on (\ref{app13}), we obtain
\begin{equation}
\begin{aligned}
& \frac{\partial Z}{\partial\mu}=\int{\frac{c-\mu}{\sigma^2}\Phi\left(\frac{c-m}{v}\right)}\mathcal{N}\left(c\middle|\mu,\sigma^2\right)dc =\frac{\partial}{\partial\mu}\Phi\left(z\right) \\
& \Longleftrightarrow \frac{1}{\sigma^2}\int c\Phi\left(\frac{c-m}{v}\right)\mathcal{N}\left(c\middle|\mu,\sigma^2\right)dc-\frac{\mu Z}{\sigma^2} \\
& =\frac{\mathcal{N}(z)}{v\sqrt{1+\sigma^2/v^2}},
\end{aligned}
\end{equation}
where $\partial\Phi\left(z\right)/\partial\mu=\mathcal{N}(z)\partial z/\partial\mu$. Combining with $\sigma^2/Z$, we get
\begin{equation}
\label{app16}
\mathbb{E}_q\left[c\right]=\mu+\frac{\sigma^2\mathcal{N}\left(z\right)}{\Phi\left(z\right)v\sqrt{1+\frac{\sigma^2}{v^2}}}.
\end{equation}
Similarly, the second moment is given by
\begin{equation}
\label{app17}
\begin{aligned}
 & \frac{\partial^2Z}{\partial\mu^2} \\
 & =\int{[\frac{c^2}{\sigma^4}-\frac{2\mu c}{\sigma^4}+\frac{\mu^2}{\sigma^4}-\frac{1}{\sigma^2}] \Phi\left(\frac{c-m}{v}\right)\mathcal{N}\left(c\middle|\mu,\sigma^2\right)} dc  \\
 & =-\frac{z\mathcal{N}(z)}{v^2+\sigma^2} \Longleftrightarrow \\
 & \mathbb{E}_q\left[c^2\right]=2\mu\mathbb{E}_q\left[c\right]-\mu^2+\sigma^2-\frac{\sigma^4z\mathcal{N}\left(z\right)}{\Phi\left(z\right)\left(v^2+\sigma^2\right)}.
\end{aligned}
\end{equation}
Combining (\ref{app16}) and (\ref{app17}), we derive
\begin{equation}
\begin{aligned}
& \mathbb{E}_q\left[{(c-\mathbb{E}_q\left[c\right])}^2\right]=\mathbb{E}_q\left[c^2\right]-\mathbb{E}_q[c]^2 \\
& =\sigma^2-\frac{\sigma^4\mathcal{N}\left(z\right)}{\left(v^2+\sigma^2\right)\Phi\left(z\right)}\left(z+\frac{\mathcal{N}\left(z\right)}{\Phi\left(z\right)}\right).
\end{aligned}
\end{equation}

Thus, \textbf{Proposition \ref{prop4}} is proven.

\section{The Proof of Proposition \ref{prop5}}
\label{appd}
We derive (\ref{19-1}), (\ref{19-2}), and (\ref{19-3}) based on (\ref{app4}), (\ref{app5}), and (\ref{app6}). Therefore, \textbf{Proposition \ref{prop5}} is proven.

\section{The Proof of Proposition \ref{prop6}}
\label{appe}
The approximate mean of $f_\ast$ is given by
\begin{equation}
\begin{aligned}
& \mathbb{E}_q\left[f_\ast|C,\bm{y},\bm{c}_\ast\right]=\bm{k}_\ast^TK^{-1}\bm{\mu} \\
& =\bm{k}_\ast^TK^{-1}\left(K^{-1}+{\widetilde{\Sigma}}^{-1}\right)^{-1}{\widetilde{\Sigma}}^{-1}\widetilde{\bm{\mu}} \\
& =\bm{k}_\ast^T\left(K+\widetilde{\Sigma}\right)^{-1}\widetilde{\bm{\mu}}.
\end{aligned}
\end{equation}

Under the Gaussian approximation, the variance of $f_\ast|(C,\bm{y})$ is given by:
\begin{equation}
\begin{aligned}
& \mathbb{V}_q\left[f_\ast\middle| C,\bm{y},\bm{c}_\ast\right] = \mathbb{E}_{p(f_\ast|C,\bm{c}_\ast,\bm{f})} {f_\ast-\mathbb{E}[f_\ast|C,\bm{c}_\ast,\bm{f}]}^2 \\
& =k\left(\bm{c}_\ast,\bm{c}_\ast\right)-\bm{k}_\ast^TK^{-1}\bm{k}_\ast+\bm{k}_\ast^TK^{-1}\left(K^{-1}+\widetilde{\Sigma}\right)^{-1}K^{-1}\bm{k}_\ast \\
& =k\left(\bm{c}_\ast,\bm{c}_\ast\right)-\bm{k}_\ast^T\left(K^{-1}+\widetilde{\Sigma}\right)^{-1}\bm{k}_\ast.
\end{aligned}
\end{equation}

Next, we get:
\begin{equation}
\begin{aligned}
& q\left(y_\ast\middle| C,\bm{y},\bm{c}_\ast\right)=\mathbb{E}_q\left[\pi_\ast|C,\bm{y},\bm{c}_\ast\right] \\
& =\int\Phi\left(f_\ast\right)q\left(f_\ast\middle| C,\bm{y},\bm{c}_\ast\right)df_\ast.
\end{aligned}
\end{equation}

Using (\ref{app11}), we derive:
\begin{equation}
\label{app22}
\begin{aligned}
& q\left(y_\ast\middle| C,\bm{y},\bm{c}_\ast\right) \\
& =\Phi\left(\frac{\bm{k}_\ast^T\left(K+\widetilde{\Sigma}\right)^{-1}\widetilde{\bm{\mu}}}{\sqrt{1+k\left(\bm{c}_\ast,\bm{c}_\ast\right)-\bm{k}_\ast^T\left(K+\widetilde{\Sigma}\right)^{-1}\bm{k}_\ast}}\right).
\end{aligned}
\end{equation}

By combining (\ref{13}) and (\ref{app22}), \textbf{Proposition \ref{prop6}} is proven.

\section{The Proof of Proposition \ref{prop7}}
\label{appf}
Considering that $f_s$ and $f_\ast$, $y_s$ and $y_\ast$ are conditionally independent, $p\left(y_s,y_\ast\middle|\bm{c}_s,\bm{c}_\ast\right)$ can be represented as
\begin{equation}
\begin{aligned}
& p\left(y_s=1,y_\ast=1\middle|\bm{c}_s,\bm{c}_\ast\right) \\
& =\iint{\Phi\left(f_s\right)\Phi\left(f_\ast\right)\phi\left(f_s,f_\ast\middle|\mu_{s\ast},\Sigma_{s\ast}\right)}df_sdf_\ast \\
& =\iint{\Phi\left(f_\ast\right)\phi\left(f_\ast\middle|{\widetilde{\mu}}_\ast\left(f_s\right),{\widetilde{\sigma}}_{\ast\ast}\right)df_\ast\Phi\left(f_s\right)}\phi\left(f_s\middle|\mu_s,\sigma_{ss}\right)df_s \\
& =\int\Phi\left(\frac{{\widetilde{\mu}}_\ast\left(f_s\right)}{\sqrt{{\widetilde{\sigma}}_{\ast\ast}+1}}\right)\Phi\left(f_s\right)\phi\left(f_s\middle|\mu_s,\sigma_{ss}\right)df_s.
\end{aligned}
\end{equation}

Hence, \textbf{Proposition \ref{prop7}} is proved.

\section{The Proof of Lemma \ref{lem}}
\label{appg}
$P_{md}$ and $P_{fa}$ are represented as
\begin{equation}
\label{pmd}
P_{md}=\frac{FL}{FL+TA}
\end{equation}
and
\begin{equation}
\label{pfa}
P_{fa}=\frac{FA}{FA+TA}.
\end{equation}

$R_e$ is denoted as
\begin{equation}
\label{Re}
\begin{aligned}
& R_e=\frac{1}{N_a}\sum_{n=1}^{N_a}\mathbb{I}\left(\bm{L}_n \neq \bm{Y}_n\right) \\
& =\displaystyle\frac{FA+FL}{TL+TA+FL+FA} \\
& =\displaystyle\frac{1}{\displaystyle\frac{TL+TA+FL+FA}{FA+FL}} \\
& =\displaystyle\frac{1}{1+\displaystyle\frac{\displaystyle\frac{TL}{TA}+1}{\displaystyle\frac{FA}{TA}+\displaystyle\frac{FL}{TA}}}
\end{aligned}
\end{equation}

By combining (\ref{pmd}), (\ref{pfa}), and (\ref{Re}), \textbf{Lemma \ref{lem}} is proved.

\bibliography{ref}

\begin{thebibliography}{10}
\providecommand{\url}[1]{#1}
\csname url@samestyle\endcsname
\providecommand{\newblock}{\relax}
\providecommand{\bibinfo}[2]{#2}
\providecommand{\BIBentrySTDinterwordspacing}{\spaceskip=0pt\relax}
\providecommand{\BIBentryALTinterwordstretchfactor}{4}
\providecommand{\BIBentryALTinterwordspacing}{\spaceskip=\fontdimen2\font plus
\BIBentryALTinterwordstretchfactor\fontdimen3\font minus
  \fontdimen4\font\relax}
\providecommand{\BIBforeignlanguage}[2]{{%
\expandafter\ifx\csname l@#1\endcsname\relax
\typeout{** WARNING: IEEEtran.bst: No hyphenation pattern has been}%
\typeout{** loaded for the language `#1'. Using the pattern for}%
\typeout{** the default language instead.}%
\else
\language=\csname l@#1\endcsname
\fi
#2}}
\providecommand{\BIBdecl}{\relax}
\BIBdecl

\bibitem{zeng2024tutorial}
Y.~Zeng, J.~Chen, J.~Xu, D.~Wu, X.~Xu, S.~Jin, X.~Gao, D.~Gesbert, S.~Cui, and
  R.~Zhang, ``A tutorial on environment-aware communications via channel
  knowledge map for 6g,'' \emph{IEEE Commun. Surv. Tutor.}, 2024.

\bibitem{dhiman2024smose}
G.~Dhiman and N.~S. Alghamdi, ``Smose: Artificial intelligence-based smart city
  framework using multi-objective and iot approach for consumer electronics
  application,'' \emph{IEEE Transactions on Consumer Electronics}, vol.~70,
  no.~1, pp. 3848--3855, 2024.

\bibitem{trevlakis2023localization}
S.~E. Trevlakis, A.-A.~A. Boulogeorgos, D.~Pliatsios, J.~Querol, K.~Ntontin,
  P.~Sarigiannidis, S.~Chatzinotas, and M.~Di~Renzo, ``Localization as a key
  enabler of 6g wireless systems: A comprehensive survey and an outlook,''
  \emph{IEEE Open J. Commun. Soc.}, 2023.

\bibitem{xia2021multiple}
S.~Xia, X.~Tao, N.~Li, S.~Wang, T.~Sui, H.~Wu, J.~Xu, and Z.~Han, ``Multiple
  correlated attributes based physical layer authentication in wireless
  networks,'' \emph{IEEE Trans. Veh. Technol.}, vol.~70, no.~2, pp. 1673--1687,
  2021.

\bibitem{garzon2022decentralized}
S.~R. Garzon, H.~Yildiz, and A.~K{\"u}pper, ``Decentralized identifiers and
  self-sovereign identity in 6g,'' \emph{IEEE Netw.}, vol.~36, no.~4, pp.
  142--148, 2022.

\bibitem{3gpp2020security}
3GPP, ``Security architecture and procedures for 5g system, version 17.0.0,''
  2020.

\bibitem{xie2022security}
N.~Xie, J.~Zhang, and Q.~Zhang, ``Security provided by the physical layer in
  wireless communications,'' \emph{IEEE Netw.}, 2022.

\bibitem{abdulqadder2022sliceblock}
I.~H. Abdulqadder and S.~Zhou, ``Sliceblock: context-aware authentication
  handover and secure network slicing using dag-blockchain in edge-assisted
  sdn/nfv-6g environment,'' \emph{IEEE Internet Things J.}, vol.~9, no.~18, pp.
  18\,079--18\,097, 2022.

\bibitem{fang2023collaborative}
H.~Fang, Z.~Xiao, X.~Wang, L.~Xu, and L.~Hanzo, ``Collaborative authentication
  for 6g networks: An edge intelligence based autonomous approach,'' \emph{IEEE
  Trans. Inf. Forensic Secur.}, vol.~18, pp. 2091--2103, 2023.

\bibitem{jin2021introduction}
L.~Jin, X.~Hu, Y.~Lou, Z.~Zhong, X.~Sun, H.~Wang, and J.~Wu, ``Introduction to
  wireless endogenous security and safety: Problems, attributes, structures and
  functions,'' \emph{China Commun.}, vol.~18, no.~9, pp. 88--99, 2021.

\bibitem{jing2023multi}
T.~Jing, H.~Huang, Q.~Gao, Y.~Wu, Y.~Huo, and Y.~Wang, ``Multi-user physical
  layer authentication based on csi using resnet in mobile iiot,'' \emph{IEEE
  Trans. Inf. Forensics Secur.}, 2023.

\bibitem{meng2023multiuser}
R.~Meng, X.~Xu, H.~Sun, H.~Zhao, B.~Wang, S.~Han, and P.~Zhang, ``Multiuser
  physical-layer authentication based on latent perturbed neural networks for
  industrial internet of things,'' \emph{IEEE Internet Things J.}, vol.~10,
  no.~1, pp. 637--652, 2023.

\bibitem{oligeri2022past}
G.~Oligeri, S.~Sciancalepore, S.~Raponi, and R.~Di~Pietro, ``Past-ai:
  Physical-layer authentication of satellite transmitters via deep learning,''
  \emph{IEEE Trans. Inf. Forensics Secur.}, vol.~18, pp. 274--289, 2022.

\bibitem{xie2020survey}
N.~Xie, Z.~Li, and H.~Tan, ``A survey of physical-layer authentication in
  wireless communications,'' \emph{IEEE Commun. Surv. Tutor.}, vol.~23, no.~1,
  pp. 282--310, 2020.

\bibitem{nguyen2021security}
V.-L. Nguyen, P.-C. Lin, B.-C. Cheng, R.-H. Hwang, and Y.-D. Lin, ``Security
  and privacy for 6g: A survey on prospective technologies and challenges,''
  \emph{IEEE Commun. Surv. Tutor.}, vol.~23, no.~4, pp. 2384--2428, 2021.

\bibitem{xiao2008using}
L.~Xiao, L.~J. Greenstein, N.~B. Mandayam, and W.~Trappe, ``Using the physical
  layer for wireless authentication in time-variant channels,'' \emph{IEEE
  Trans. Commun.}, vol.~7, no.~7, pp. 2571--2579, 2008.

\bibitem{liu2022online}
Y.~Liu, P.~Zhang, Y.~Shen, L.~Peng, and X.~Jiang, ``Online machine
  learning-based physical layer authentication for mmwave mimo systems,''
  \emph{Ad Hoc Networks}, vol. 131, p. 102864, 2022.

\bibitem{chorti2022context}
A.~Chorti, A.~N. Barreto, S.~K{\"o}psell, M.~Zoli, M.~Chafii, P.~Sehier,
  G.~Fettweis, and H.~V. Poor, ``Context-aware security for 6g wireless: The
  role of physical layer security,'' \emph{IEEE Communications Standards
  Magazine}, vol.~6, no.~1, pp. 102--108, 2022.

\bibitem{meng2024survey}
R.~Meng, B.~Xu, X.~Xu, M.~Sun, B.~Wang, S.~Han, S.~Lv, and P.~Zhang, ``A survey
  of machine learning-based physical-layer authentication in wireless
  communications,'' \emph{Journal of Network and Computer Applications}, p.
  104085, 2024.

\bibitem{wang2024knowledge}
Q.~Wang, W.~Liang, J.~Zhang, K.~Wang, and X.~Jiang, ``Knowledge-enhanced
  physical layer authentication for mobile devices,'' \emph{IEEE Transactions
  on Consumer Electronics}, vol.~70, no.~4, pp. 7436--7448, 2024.

\bibitem{wang2021channel}
H.-M. Wang and Q.-Y. Fu, ``Channel-prediction-based one-class mobile iot device
  authentication,'' \emph{IEEE Internet Things J.}, vol.~9, no.~10, pp.
  7731--7745, 2022.

\bibitem{xie2021weighted}
F.~Xie, Z.~Pang, H.~Wen, W.~Lei, and X.~Xu, ``Weighted voting in physical layer
  authentication for industrial wireless edge networks,'' \emph{IEEE Trans.
  Ind. Inform.}, vol.~18, no.~4, pp. 2796--2806, 2021.

\bibitem{chen2021physical}
Y.~Chen, P.-H. Ho, H.~Wen, S.~Y. Chang, and S.~Real, ``On physical-layer
  authentication via online transfer learning,'' \emph{IEEE Internet Things
  J.}, vol.~9, no.~2, pp. 1374--1385, 2021.

\bibitem{xiao2016phy}
L.~Xiao, Y.~Li, G.~Han, G.~Liu, and W.~Zhuang, ``Phy-layer spoofing detection
  with reinforcement learning in wireless networks,'' \emph{IEEE Trans. Veh.
  Technol.}, vol.~65, no.~12, pp. 10\,037--10\,047, 2016.

\bibitem{recommendation2023framework}
I.~RECOMMENDATION, ``Framework and overall objectives of the future development
  of imt for 2030 and beyond,'' tech. rep., International Telecommunication
  Union (ITU) Recommendation (ITU-R), Tech. Rep., 2023.

\bibitem{fang2018learning}
H.~Fang, X.~Wang, and L.~Hanzo, ``Learning-aided physical layer authentication
  as an intelligent process,'' \emph{IEEE Trans. Commun.}, vol.~67, no.~3, pp.
  2260--2273, 2018.

\bibitem{zhang2020physical}
P.~Zhang, Y.~Shen, X.~Jiang, and B.~Wu, ``Physical layer authentication jointly
  utilizing channel and phase noise in mimo systems,'' \emph{IEEE Trans.
  Commun.}, vol.~68, no.~4, pp. 2446--2458, 2020.

\bibitem{senigagliesi2022authentication}
L.~Senigagliesi, M.~Baldi, and E.~Gambi, ``Authentication at the physical layer
  with cooperative communications and machine learning,'' in \emph{2022 Joint
  European Conference on Networks and Communications \& 6G Summit (EuCNC/6G
  Summit)}.\hskip 1em plus 0.5em minus 0.4em\relax IEEE, 2022, pp. 71--76.

\bibitem{xiao2017phy}
L.~Xiao, X.~Wan, and Z.~Han, ``Phy-layer authentication with multiple landmarks
  with reduced overhead,'' \emph{IEEE Trans. Commun.}, vol.~17, no.~3, pp.
  1676--1687, 2017.

\bibitem{meng2024multiobservation}
R.~Meng, X.~Xu, H.~Zhao, B.~Wang, G.~Li, B.~Xu, and P.~Zhang,
  ``Multiobservation-multichannel-attribute-based multiuser authentication for
  industrial wireless edge networks,'' \emph{IEEE Trans. Ind. Informat.},
  vol.~20, no.~2, pp. 2097--2108, 2024.

\bibitem{meng2023physical}
R.~Meng, X.~Xu, B.~Wang, H.~Sun, S.~Xia, S.~Han, and P.~Zhang, ``Physical-layer
  authentication based on hierarchical variational autoencoder for industrial
  internet of things,'' \emph{IEEE Internet Things J.}, vol.~10, no.~3, pp.
  2528--2544, 2023.

\bibitem{chen2023optimal}
S.~Chen, Y.~Ji, Y.~Jiang, W.~Duan, J.~Choi, G.~Zhang, and P.-H. Ho, ``Optimal
  ris allocations for pls with uncertain jammer and eavesdropper,'' \emph{IEEE
  Transactions on Consumer Electronics}, vol.~69, no.~4, pp. 927--936, 2023.

\bibitem{chen2022accurate}
M.~Z. Chen, W.~Tang, J.~Y. Dai, J.~C. Ke, L.~Zhang, C.~Zhang, J.~Yang, L.~Li,
  Q.~Cheng, S.~Jin \emph{et~al.}, ``Accurate and broadband manipulations of
  harmonic amplitudes and phases to reach 256 qam millimeter-wave wireless
  communications by time-domain digital coding metasurface,'' \emph{Natl. Sci.
  Rev.}, vol.~9, no.~1, p. nwab134, 2022.

\bibitem{gao2024physical}
N.~Gao, Y.~Han, N.~Li, S.~Jin, and M.~Matthaiou, ``When physical layer key
  generation meets ris: Opportunities, challenges, and road ahead,'' \emph{IEEE
  Wirel. Commun.}, 2024.

\bibitem{williams2006gaussian}
C.~K. Williams and C.~E. Rasmussen, \emph{Gaussian processes for machine
  learning}.\hskip 1em plus 0.5em minus 0.4em\relax MIT press Cambridge, MA,
  2006, vol.~2, no.~3.

\bibitem{zhu2022intelligent}
Y.~Zhu, B.~Mao, and N.~Kato, ``Intelligent reflecting surface in 6g vehicular
  communications: A survey,'' \emph{IEEE Open J. Veh. Technol.}, vol.~3, pp.
  266--277, 2022.

\bibitem{settles2009active}
B.~Settles, \emph{Active learning literature survey}.\hskip 1em plus 0.5em
  minus 0.4em\relax University of Wisconsin-Madison Department of Computer
  Sciences, 2009.

\bibitem{tomasin2022challenge}
S.~Tomasin, H.~Zhang, A.~Chorti, and H.~V. Poor, ``Challenge-response physical
  layer authentication over partially controllable channels,'' \emph{IEEE
  Commun. Mag.}, vol.~60, no.~12, pp. 138--144, 2022.

\bibitem{wang2022wireless}
Y.~Wang, H.~Lu, D.~Zhao, Y.~Deng, and A.~Nallanathan, ``Wireless communication
  in the presence of illegal reconfigurable intelligent surface: Signal leakage
  and interference attack,'' \emph{IEEE Wirel. Commun.}, vol.~29, no.~3, pp.
  131--138, 2022.

\bibitem{zheng2022survey}
B.~Zheng, C.~You, W.~Mei, and R.~Zhang, ``A survey on channel estimation and
  practical passive beamforming design for intelligent reflecting surface aided
  wireless communications,'' \emph{IEEE Commun. Surv. Tutor.}, vol.~24, no.~2,
  pp. 1035--1071, 2022.

\bibitem{ghosh2024ee}
S.~Ghosh, S.~P. Maity, and C.~Chakraborty, ``On ee maximization in d2d-crn with
  eavesdropping using lstm based channel estimation,'' \emph{IEEE Transactions
  on Consumer Electronics}, vol.~70, no.~1, pp. 3906--3913, 2024.

\bibitem{dalton2013optimal}
L.~A. Dalton and E.~R. Dougherty, ``Optimal classifiers with minimum expected
  error within a bayesian framework—part i: Discrete and gaussian models,''
  \emph{Pattern Recognit.}, vol.~46, no.~5, pp. 1301--1314, 2013.

\bibitem{yoon2013quantifying}
B.-J. Yoon, X.~Qian, and E.~R. Dougherty, ``Quantifying the objective cost of
  uncertainty in complex dynamical systems,'' \emph{IEEE Trans. Signal
  Process.}, vol.~61, no.~9, pp. 2256--2266, 2013.

\bibitem{zhao2021efficient}
G.~Zhao, E.~Dougherty, B.-J. Yoon, F.~Alexander, and X.~Qian, ``Efficient
  active learning for gaussian process classification by error reduction,''
  \emph{Advances in Neural Information Processing Systems}, vol.~34, pp.
  9734--9746, 2021.

\bibitem{roy2001toward}
N.~Roy and A.~McCallum, ``Toward optimal active learning through monte carlo
  estimation of error reduction,'' \emph{ICML, Williamstown}, vol.~2, pp.
  441--448, 2001.

\bibitem{zhao2021bayesian}
G.~Zhao, E.~Dougherty, B.-J. Yoon, F.~J. Alexander, and X.~Qian, ``Bayesian
  active learning by soft mean objective cost of uncertainty,'' in
  \emph{International Conference on Artificial Intelligence and
  Statistics}.\hskip 1em plus 0.5em minus 0.4em\relax PMLR, 2021, pp.
  3970--3978.

\bibitem{3gpp2022study}
3GPP, ``Study on channel model for frequencies from 0.5 to 100 ghz, version
  17.0.0,'' 2022.

\bibitem{hu2021reconfigurable}
X.~Hu, C.~Masouros, and K.-K. Wong, ``Reconfigurable intelligent surface aided
  mobile edge computing: From optimization-based to location-only
  learning-based solutions,'' \emph{IEEE Trans. Commun.}, vol.~69, no.~6, pp.
  3709--3725, 2021.

\bibitem{sebastiani2000maximum}
P.~Sebastiani and H.~P. Wynn, ``Maximum entropy sampling and optimal bayesian
  experimental design,'' \emph{Journal of the Royal Statistical Society: Series
  B (Statistical Methodology)}, vol.~62, no.~1, pp. 145--157, 2000.

\bibitem{houlsby2011bayesian}
N.~Houlsby, F.~Husz{\'a}r, Z.~Ghahramani, and M.~Lengyel, ``Bayesian active
  learning for classification and preference learning,'' \emph{arXiv preprint
  arXiv:1112.5745}, 2011.

\end{thebibliography}
\bibliographystyle{IEEEtran}

\vfill

\end{document}